\documentclass[aps,prd,superscriptaddress,preprint,tightenlines,floatfix,nofootinbib]{revtex4}

\usepackage{graphicx} 
\usepackage{dcolumn}  
\usepackage{bm}       

\begin{document}

\newcommand{\PM}{$\pm$}

\preprint{CLNS 08/2018}       
\preprint{CLEO 08-02}         

\title{\boldmath Dalitz plot analysis of the $D^+\to K^-\pi^+\pi^+$ decay}

\author{G.~Bonvicini}
\author{D.~Cinabro}
\author{M.~Dubrovin}
\author{A.~Lincoln}
\affiliation{Wayne State University, Detroit, Michigan 48202, USA}
\author{P.~Naik}
\author{J.~Rademacker}
\affiliation{University of Bristol, Bristol BS8 1TL, United Kingdom}
\author{D.~M.~Asner}
\author{K.~W.~Edwards}
\author{J.~Reed}
\affiliation{Carleton University, Ottawa, Ontario, Canada K1S 5B6}
\author{R.~A.~Briere}
\author{T.~Ferguson}
\author{G.~Tatishvili}
\author{H.~Vogel}
\author{M.~E.~Watkins}
\affiliation{Carnegie Mellon University, Pittsburgh, Pennsylvania 15213, USA}
% ...
\collaboration{CLEO Collaboration}
\noaffiliation

\date{September 12, 2008}

\begin{abstract} 
We perform a Dalitz plot analysis of $D^+ \to K^- \pi^+ \pi^+$ decay
with the CLEO-c data set of 572~pb$^{-1}$ of $e^+e^-$ collisions
accumulated at the $\psi(3770)$.  This corresponds to
$1.6 \times 10^6$ $D^+D^-$ pairs from which we select 140793 candidate events
with a small background of 1.1\%.
We compare our results with previous measurements
using the isobar model.
We modify the isobar model with an improved description of
some of the contributing resonances, and get better agreement with our data.
We also consider a quasi-model-independent approach and
measure the magnitude and phase of the contributing $K\pi$ $S$ wave in the range
of invariant masses from the threshold to the maximum in this decay.
This gives an improved descriptions of our data over the isobar model.
Finally we allow for an isospin-two $\pi^+ \pi^+$ $S$ wave contribution,
and find that adding this to both the isobar model and the quasi-model-independent
approach gives the best description of our data.
\end{abstract}

     
\maketitle

\section{Introduction}
\label{sec:introduction}

In comparison to other $D^+$ decay modes, 
the $D^+ \to K^-\pi^+\pi^+$ decay is unique in many aspects.
The large branching fraction, 
${\cal B}(D^+ \to K^- \pi^+ \pi^+)=(9.51 \pm 0.34)\%$~\cite{PDG-2006}, 
for this Cabibbo favored mode 
makes it the usual choice for normalization of other $D^+$-meson decay rates.
Understanding its peculiar intermediate substructure will be beneficial.
The only obvious contribution to this decay, observed in the $K\pi$-mass spectrum,
is $K^*(892)^0\pi^+$, which comprises merely 12\% of the total rate~\cite{PDG-2006}.
A large contribution of over 60\% from a $K\pi$ $S$ wave intermediate state 
has been observed in earlier experiments, including
MARK III \cite{MARK-III-1987},
NA14     \cite{NA14-1991},
E691     \cite{E691-1993},
E687     \cite{E687-1994}, and
E791     \cite{E791-2002,E791_Kpipi},
where the $D^+ \to K^-\pi^+\pi^+$ decay has been studied
with the Dalitz plot technique~\cite{Dalitz}.
Hence, the $D^+ \to K^-\pi^+\pi^+$ decay is a good laboratory to study
$K\pi$ $S$ wave dynamics. 

The previous analysis by E791 \cite{E791-2002} achieved good
agreement with their data by including
a low-mass $K^-\pi^+$ scalar resonance $\kappa$ that significantly
redistributed
all fit fractions (FF) observed by earlier experiments. 
This particular model, even though it is based on the largest data
set, greatly disagrees with previous analyses and has been excluded from
the average given by the Particle Data Group (PDG)~\cite{PDG-2006}.

There has been significant theoretical interest in this decay, sparked
by the large, low-mass $K\pi$ $S$ wave contribution.
In Refs.\cite{Bugg-2005,Bugg-2006,Oller-2005}
the authors reanalyze the E791 data with their own models.
E791 later reinterpreted their own data
with a model-independent partial wave analysis \cite{E791_Kpipi},
and we apply this in our analysis with minor modifications.

The two identical pions in the final state should obey Bose symmetry.
Assuming that the three-body decay is dominated by
two-body intermediate states, 
there would be two identical $K^-\pi^+$ waves interfering with each other.
This two-fold symmetry significantly reduces the degrees of freedom 
in the regular Dalitz plot analysis
and allows the application of a model-independent
partial wave analysis \cite{E791_Kpipi}.
We would also expect a small contribution from the
isospin-two $\pi^+\pi^+$ $S$ wave,
which exhibits nontrivial dynamics as  observed 
in scattering experiments~\cite{Hoodland}.

The data used in this analysis were accumulated with the CLEO-c detector~\cite{CLEO-c}. 
Our event sample is based on 572~pb$^{-1}$ of $e^+e^-$ collisions at $\sqrt{s}\approx 3774$~MeV,
produced by the Cornell Electron Storage Ring (CESR).
This sample
corresponds to the production of  $1.6 \times 10^6$ $D^+D^-$ pairs in the process
$e^+e^- \to \psi(3770) \to D^+D^-$.
We select 140793 $D^+ \to K^-\pi^+\pi^+$ candidates 
for the Dalitz plot analysis
(charge conjugation is implied throughout this paper).  
Our sample is very clean with a background fraction of about 1.1\% 
and is 9 times larger than the data set used by E791.
The invariant mass resolution in this three-track $D$-meson decay is
very good; we estimate it is better than 5~MeV/$c^2$ in most cases. 
It is improved by a kinematic fit requiring a 
three-track common vertex with the $D$-meson mass constraint.
Our kinematic conditions are similar to those of
MARK III, where $D$ mesons are produced with small momentum.

In Sec.~\ref{sec:detector} we briefly discuss CLEO-c experimental techniques,
giving the event selection for the Dalitz plot analysis,
the general fit method and methods to parametrize the signal
efficiency and background distribution across the Dalitz plot.
The formalism we use for the amplitude parametrization in this analysis 
is described in Sec.~\ref{sec:formalism}.  
In Sec.~\ref{sec:isobar_model}
we compare our results with the best previous measurements by E791 \cite{E791-2002}
and try to improve the isobar model in order to get a better description of our data.
Finally, we apply a quasi-model-independent partial wave analysis, 
following Ref.~\cite{E791_Kpipi}
and measure the partial waves contributing to this decay in Sec.~\ref{sec:QMIPWA_I2}.
Systematic studies and cross-checks are considered in Sec.~\ref{sec:systematic_cross_checks}.
We discuss results and outstanding issues of this analysis
in Sec.~\ref{sec:discussion} 
and summarize our results in Sec.~\ref{sec:summary}.
In the appendix 
we discuss the kinematic variables 
and angular distributions used in this analysis.

\section{Detector and experimental technique}
\label{sec:detector}
\subsection{Detector} 

CLEO-c is a general purpose detector which includes a tracking system
for measuring momenta and specific ionization of charged particles,
a ring imaging Cherenkov detector to aid particle identification,
and a CsI calorimeter for detection of electromagnetic showers.
These components are immersed in a magnetic field of 1~T,
provided by a superconducting solenoid, and surrounded by a muon detector.
The CLEO-c detector is described in detail elsewhere~\cite{CLEO-c}.

\subsection{Event Reconstruction}
\label{sec:event_reconstrucion}

We reconstruct the $D^+ \to K^-\pi^+\pi^+$ decay
using three tracks measured in the tracking system.
Charged tracks satisfy standard goodness of fit quality requirements~\cite{HadronicBF}.
Pion and kaon candidates are required to have specific ionization
$dE/dx$ in the main drift chamber within 4 standard
deviations of the expected value at the measured momentum.

\begin{figure}[!htb]
  \includegraphics[width=160mm]{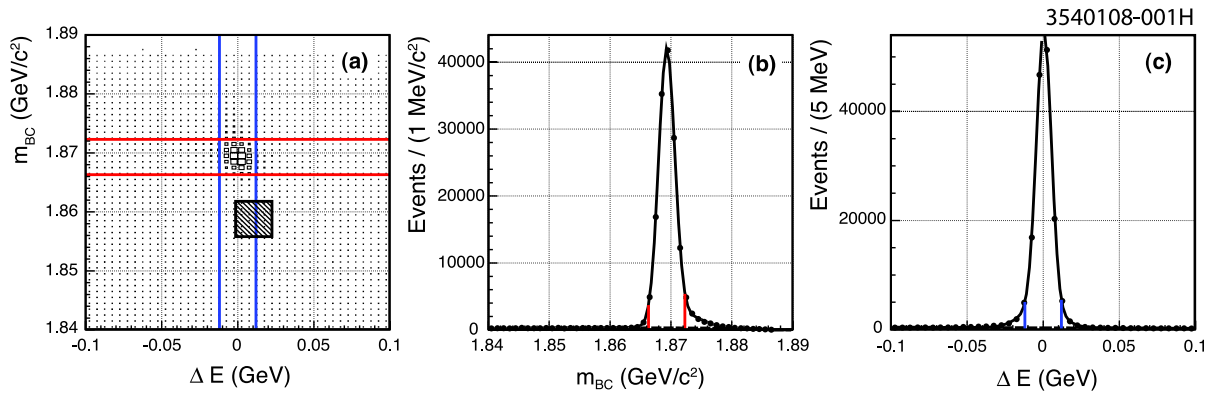}
  \caption{\label{fig:selection_1} Event selection, 
    (a) The $m_{\rm BC}$ and $|\Delta E|$ two-dimensional distribution for
    data and the projections on (b) $m_{\rm BC}$ and (c) $\Delta E$. The
    $m_{\rm BC}$ and $\Delta E$ signal regions, defined in the text, are
    shown as the bands in the figures. In (a), the ``signal box''
    is indicated as the crossing area of the two bands while the ``sideband
    box,'' defined in the text, is indicated as the shaded rectangle. Each
    projection is made with the events in the signal region of the other
    kinematic variable; the fit curve, described in the text, is also shown.}
\end{figure}

In order to select $D^+ \to K^-\pi^+\pi^+$ decays,
we use two kinematic variables,
\begin{equation}
      \Delta E = E_D - E_{\mathrm{\rm beam}},
\end{equation}
\begin{equation}
      m_{\rm BC} = \sqrt{E^2_{\rm beam}-{\rm P}^2_D},
\end{equation}
where $E_{\rm beam}$ is the beam energy,
and $E_D$ and ${\rm P}_D$ are the energy and the magnitude, respectively, of the momentum of
the reconstructed $D^+$ candidate.
The $m_{\rm BC}$ and $|\Delta E|$ two-dimensional distribution and
the projections for data are shown in Fig.~\ref{fig:selection_1}. The resolutions in
$\Delta E$ and $m_{\rm BC}$ are represented as $\sigma(\Delta E) = 6$
MeV and $\sigma(m_{\rm BC})= 1.5$ MeV/$c^2$, respectively; fits with a
Gaussian function to the $\Delta E$ and $m_{\rm BC}$ peaks evaluate the
resolutions to be $5.539\pm 0.014$ MeV and $1.410\pm 0.013$ MeV/$c^2$,
respectively. We require the events to fall in the ``signal box" that
is the overlap region of the $\Delta E$ and $m_{\rm BC}$ signal regions
defined as $|\Delta E| < 2\sigma(\Delta E)$ and $|m_{\rm BC} - m_D| < 2
\sigma(m_{\rm BC})$, respectively.
In the case of multiple $D$-meson candidates per event we select the one
with the smallest $|\Delta E|$ value.

\begin{figure}[!htb]
  \includegraphics[width=160mm]{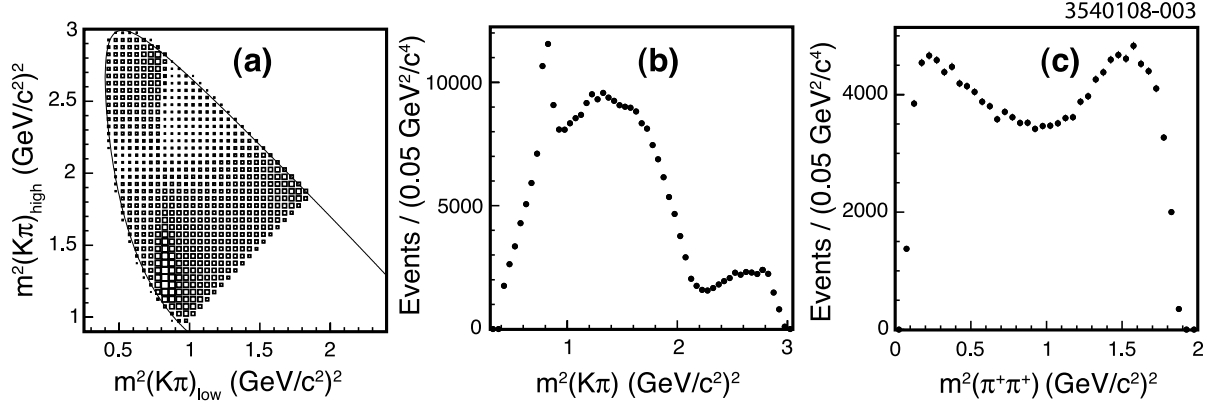}
  \caption{\label{fig:DP_data} (a) Dalitz plot for data and their projections
                                on (b) $m^2(K\pi)$ (two entries per event), and 
                                   (c) $m^2(\pi\pi)$ variables.} 
\end{figure}

\begin{figure}[!htb]
  \includegraphics[width=160mm]{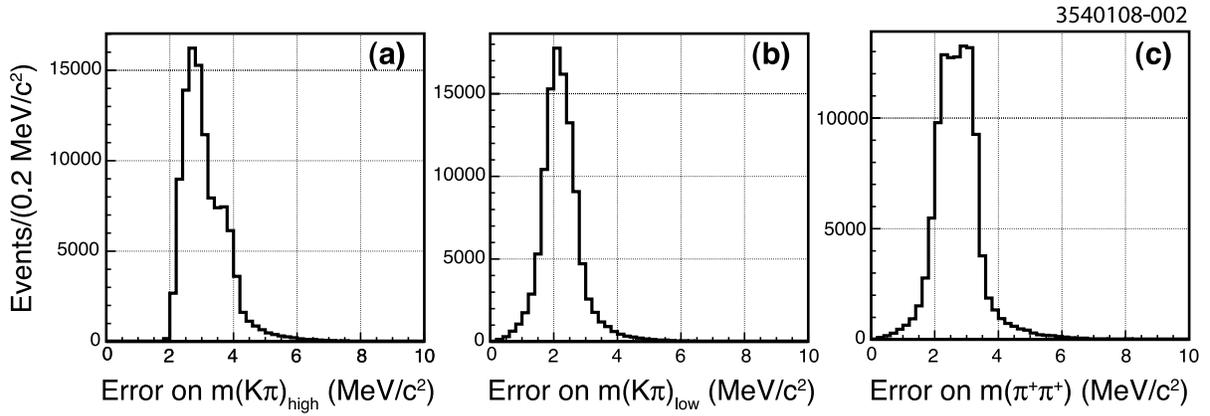}
  \caption{\label{fig:inv_mass_resolutions} Invariant mass resolutions before the
                  kinematic fit for 
                  (a) $m(K\pi)_{\rm high}$, (b) $m(K\pi)_{\rm low}$, and (c) $m(\pi\pi)$.} 
\end{figure}

The $K^-\pi^+\pi^+$ final state has two identical $\pi^+$ mesons.
To account for this symmetry
we analyze events on the Dalitz plot by choosing
$x = m^2(K^-\pi^+)_{\rm low}$ and 
$y = m^2(K^-\pi^+)_{\rm high}$ as the
independent ($x,y$) variables. 
This choice folds all of
the data onto the top half of the kinematically allowed region,
as is shown in Fig.~\ref{fig:DP_data}(a). 
The third variable $z = m^2(\pi^+\pi^+)$ is dependent on $x$ and $y$
through energy and momentum conservation.  
The invariant mass resolutions, propagated from the track error matrices,
are shown in Fig.~\ref{fig:inv_mass_resolutions}, 
and in 95\% of cases are better than 5~MeV/$c^2$.
We use a kinematic fit to all 3-track candidates which enforces a
common vertex and $D^+$ mass~\cite{PDG-2006} constraint.
We require that all events pass the kinematic fit successfully but 
do not restrict their $\chi^2$. 
The kinematic-fit-corrected 4-momenta of all 3 particles are used to 
calculate invariant masses for further Dalitz plot analysis.
Within its finite accuracy, the kinematic fit improves the $K^-\pi^+\pi^+$ invariant mass 
resolution by 2 orders of magnitude. Proportional improvement is
expected for all two-body invariant mass resolutions.

After all requirements,
we select 140793 events for the Dalitz plot analysis.
The signal fraction in this sample $f_{\rm sig}$ is estimated to be
$(98.917\pm 0.013)\%$ from the fit to the $m_{\rm BC}$ distribution,
shown in Fig.~\ref{fig:selection_1}(b). In this fit the signal 
and background shapes are described by the double-Gaussian and
ARGUS~\cite{ARGUS_fun} functions, respectively, with all parameters free. 
This value of $f_{\rm sig}$ is consistent with one obtained from 
the fit to $\Delta E$ distribution, shown in Fig.~\ref{fig:selection_1}(c).
In most fits to the Dalitz plot
we use the fixed value of the signal fraction.
Figure~\ref{fig:DP_data} shows the Dalitz plot data and
two projections onto the $m^2(K\pi)$ axes
[two entries per event for $m^2(K\pi)_{\rm low}$ and $m^2(K\pi)_{\rm high}$, respectively]
and $m^2(\pi\pi)$.
Besides the clear $K^*(892)$ signal, no other narrow features are obvious.  
The strong left-right asymmetry of the $K^*(892)$ population density 
on the Dalitz plot is evidence of the interference between $P$ and $S$ waves.
There are broad structures, including a
  peak at $m^2(K\pi)$   around 1.3~GeV$^2$/$c^4$, 
a  dip at $m^2(K\pi)$   around 2.25~GeV$^2$/$c^4$, a 
   dip at $m^2(\pi\pi)$ around 1~GeV$^2$/$c^4$, and 
a peak at $m^2(\pi\pi)$ around 1.6~GeV$^2$/$c^4$, 
that do not obviously correspond with known resonances or their reflections
from other axes. These structures also
do not correspond to a flat phase space distribution for nonresonant
decays, since our efficiency is essentially flat across the Dalitz plot,
and the background is very low.  Thus, we are compelled to consider
$K\pi$ strong interaction dynamics to explain the Dalitz plot.

\subsection{Fit method}
\label{sec:fit_method}

In order to describe the event density distribution on the Dalitz plot we use 
a probability density function (p.d.f.), ${\cal P}(x,y)$,
which depends on the event sample being fit:
\begin{equation}
\label{eqn:PDF}
{\cal P}(x,y) = \left\{
           \begin{array}{ll}
                  {\mathcal N}_{\varepsilon} \varepsilon(x,y) & {\rm for~efficiency,} \\
                  {\mathcal N}_B B(x,y)                       & {\rm for~background,} \\
                  f_{\rm{sig}} {\mathcal N}_S |{\mathcal M}(x,y)|^2 \varepsilon(x,y)
                + (1-f_{\rm{sig}}) {\mathcal N}_B B(x,y)
                                                              & {\rm for~signal~with~background,}
            \end{array}
              \right.
\end{equation}
where the $\varepsilon(x,y)$ and $B(x,y)$ are the functions representing the 
shape of the efficiency and background, respectively, across the Dalitz plot.
The signal p.d.f.\ is proportional to the
efficiency-corrected matrix element squared $|{\mathcal M}(x,y)|^2$, 
defined in Sec.~\ref{sec:formalism},
whose fraction $f_{\rm sig}$ is introduced earlier.
The background term has a relative $(1-f_{\rm sig})$ fraction.
All p.d.f.\ components are normalized separately using the normalization integrals
over the Dalitz plot area
$1/{\mathcal N}_{\varepsilon} = \int \varepsilon(x,y) dx dy$,
$1/{\mathcal N}_B = \int B(x,y) dx dy$, and
$1/{\mathcal N}_S = \int |{\mathcal M}(x,y)|^2 \varepsilon(x,y) dx dy$,
which provides the overall p.d.f.\ normalization,
$\int {\cal P}(x,y) dx dy =1$.
The p.d.f.\ free parameters are optimized with
an maximum likelihood fit that minimizes
the sum over $N$ events:
\begin{equation}
\label{eqn:LogL}
        \mathcal{L} = -2\sum_{n=1}^{N} \log {\cal P}(x_n,y_n).
\end{equation}
To estimate the fit quality we use Pearson's statistics for adaptive bins, 
similar to our previous analysis~\cite{D3pi}.

\subsection{Efficiency Parametrization}
\label{sec:efficiency}

To determine the efficiency we use a signal Monte Carlo (MC) 
\cite{EVTGEN} simulation
where one of the charged $D$ mesons decays in the signal mode
uniformly in phase space, while the other $D$ meson decays
in all known modes with relevant branching fractions.
These underlying events are input to the CLEO-c detector
simulation and processed with the regular reconstruction package.
The MC-generated events are required to pass the same selection requirements as data
selected in the signal box, as shown in Fig.~\ref{fig:selection_1}(a).
In each event we consider only the signal mode side
to prevent nonuniformity of the efficiency due to the resonance 
substructure of the other side $D$ decay currently implemented in our generic simulation.
The efficiency of the $K\pi\pi$ final state selection for the Dalitz plot analysis
is estimated to be ($51.11\pm 0.07$)\% where the error is only statistical. 
This number also accounts for a correction factor, $0.984$, 
due to the nonuniform population of the data on the Dalitz plot,
\begin{equation}
\label{efficiency_correction}
f_{\rm corr} = \frac{\overline{\varepsilon}_{\rm Data}}
                    {\overline{\varepsilon}_{\rm MC}}
             = \frac{\sum_{n=1}^N \varepsilon(x_n,y_n)~ \big/~ N  }
                    {\int  \varepsilon(x,y) dx dy~ \big/~ \int  dx dy },
\end{equation}
where $\overline \varepsilon_{\rm Data}$ and $\overline \varepsilon_{\rm MC}$
are the average efficiencies for data and uniformly generated MC samples,
the function $\varepsilon(x,y)$ is an efficiency over Dalitz plot
defined later by Eqs.~(\ref{eqn:efficiency}) and
                      (\ref{eqn:threshold_factor}),
the sum runs over all $N$ events in the data sample, and the integrals are taken
over the area of the Dalitz plot.

\begin{figure}[!htb]
  \includegraphics[width=160mm]{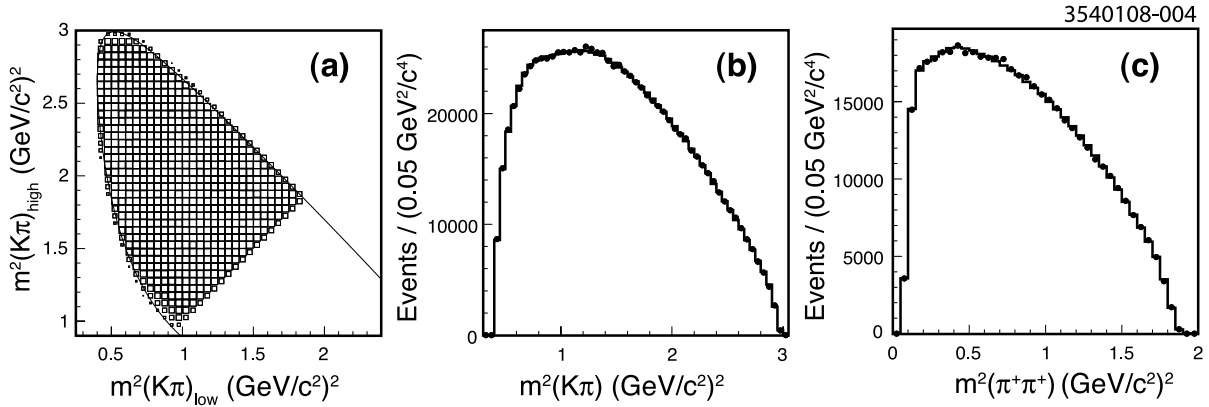}
  \caption{\label{fig:efficiency} For the efficiency shape:
                                  (a) Dalitz plot of the signal MC generated uniformly in
                                  phase space
                                  and its projections on (b) $m^2(K\pi)$ (two entries per event) and 
                                  (c) $m^2(\pi\pi)$ variables. The solid histogram is a projection of the 
                                  function described in the text which parametrizes the efficiency.  
                                  Binned results are shown, but the efficiency shape
                                  is determined with an unbinned maximum likelihood fit.  } 
\end{figure}

To parametrize the efficiency $\varepsilon (x,y)$,
we use a symmetric third-order polynomial function 
centered on the arbitrary 
point ($x_c$, $y_c$)=(1.5, 1.5)~(GeV/$c^2$)$^2$ on the Dalitz plot.
With
$\hat{x} \equiv x-x_c$ and
$\hat{y} \equiv y-y_c$,
the efficiency is the product of the polynomial function:
\begin{equation}
\label{eqn:efficiency}
\varepsilon (x,y) = T(v)[1 + E_1(\hat{x}+\hat{y}) + E_2(\hat{x}^2+\hat{y}^2) + E_3(\hat{x}^3+\hat{y}^3)
                      + E_{xy}\hat{x}\hat{y} + E_{xyn}(\hat{x}^2\hat{y}+\hat{x}\hat{y}^2)],
\end{equation}
and sinelike threshold factors for each 
Dalitz plot variable $v$  ($\equiv x, y$ or $z$):

\begin{equation}
\label{eqn:threshold_factor}
   T(v) = \left\{
\begin{array}{ll}
                 \sin( E_{{\rm th},v} |v-v_{\rm max}| ), & 
                  {\rm ~~~for~~ } 0<E_{{\rm th},v} |v-v_{\rm max}|<\pi/2, \\
                 1                                         , & 
                  {\rm ~~~for~~ } E_{{\rm th},v} |v-v_{\rm max}| \geq \pi/2, \\
\end{array} \right.
\end{equation}
where all polynomial coefficients 
$E_1$, $E_2$, $E_3$, $E_{xy}$,  $E_{xyn}$, and $E_{{\rm th},v}$ 
are the fit parameters.
Each variable $v$ has two thresholds $v_{\rm min}$ and $v_{\rm max}$.
We expect low efficiency in the regions $v \approx v_{\rm max}$ only, where
one of three particles is produced with zero momentum in the $D$-meson rest frame
and thus has a small momentum in the laboratory frame.
Figure~\ref{fig:efficiency} and
Table~\ref{tab:Efficiency}
show results of the fit to the entire signal MC sample
of $D^+ \to K^-\pi^+\pi^+$ events selected on the Dalitz plot.
The polynomial function with threshold factors well describes the efficiency
shape. 
If we consider subsamples of our signal MC, such as $D^+$ versus $D^-$,
we find that the variation of the efficiency parameters 
is small compared to their statistical uncertainties.  
In fits to data we use this efficiency shape with fixed parameters,
and variations constrained by the errors from our fit to the
signal MC are allowed as systematic checks.

\begin{table}[!htb]
\caption{\label{tab:Efficiency} Fit parameters for the efficiency from the signal MC sample.}
\begin{center}
\begin{tabular}{ l c }
\hline
\hline
Parameter       &  Value          \\
\hline
  $E_1$         &--0.0153\PM0.0090\\
  $E_2$         &--0.030\PM0.011  \\
  $E_3$         &  0.162\PM0.020  \\
  $E_{xy}$      &--0.053\PM0.019  \\
  $E_{xyn}$     &  0.673\PM0.055  \\
  $E_{{\rm th,}x} \equiv E_{{\rm th,}y}$ 
                &  4.25\PM0.23    \\
  $E_{{\rm th,}z}$  &  2.907\PM0.075  \\
\hline
Pearson $\chi^2/\nu$  
                &  649/573        \\
Probability (\%)&  1.5            \\
Events on DP    &  477978         \\
\hline
\hline
\end{tabular}
\end{center}
\end{table}

\subsection{Background Parametrization}
\label{sec:background}

A shape for the background on the Dalitz plot is estimated
using data events from a $m_{\rm BC}$ sideband region, 
shown by the hatched box in Fig.~\ref{fig:selection_1}(a).
This box is shifted in $\Delta E$ from the signal region to have
the same $K^-\pi^+\pi^+$
invariant mass range as candidates in the signal box.
We consider only events from the low-mass $m_{\rm BC}$ sideband
as the high-mass sideband has a significant contribution from 
signal events due to a ``tail'' caused by initial state radiation.
This tail is clearly seen in the $m_{\rm BC}$ distribution shown
in Fig.~\ref{fig:selection_1}(b).

The background is a small contribution that has little effect on our fits.
Nevertheless, we study the background composition using generic
MC simulation for all known modes and find the following.
The pileup of events at $m^2(K\pi)_{\rm high}\approx 2.6$~GeV$^2/c^4$
is caused by the misreconstructed $D$
decays from $D^0\overline{D}^0$ and $D^+D^-$ pairs and
the combinatorial background from the $e^+e^-\to q\overline{q}$
($q = u$, $d$ and $s$) continuum; their contributions have
the relative fractions of 62\%, 13\% and 25\%, respectively, in this
area. The dominant misreconstructed $D$ decays are $D^0\to K^-\pi^+\pi^0$, 
$a_1(1260)^+ K^-$, and $D^+\to K^-\pi^+\pi^+$, where one of
the final state pions is misreconstructed and replaced with a $\pi^+$
meson from the other $D$ decay. The shape of the background
is well reproduced in our simulation 
for events selected from either the signal or sideband box. 

\begin{figure}[!htb]
  \includegraphics[width=160mm]{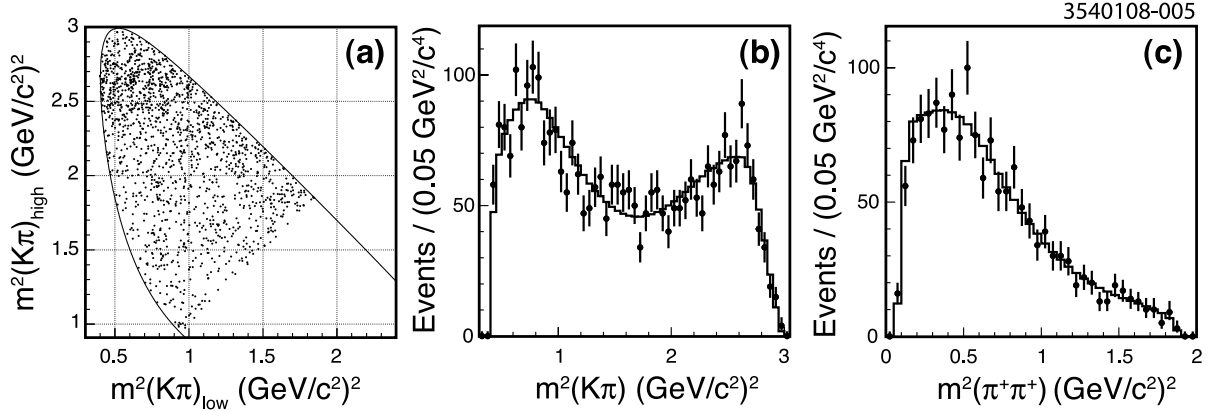}
  \caption{\label{fig:background} (a) Dalitz plot of data in the sideband box and projections on
                                  (b) $m^2(K\pi)$ (two entries per event), and 
                                  (c) $m^2(\pi\pi)$ variables.  The solid histogram shows the projection
                                  of the fit function used to parametrize the background shape 
                                  described in the text.} 
\end{figure}

To parametrize the background shape on the Dalitz plot
we employ a function similar to that used for the efficiency, shown in 
Eqs.~(\ref{eqn:efficiency}) and (\ref{eqn:threshold_factor}).
Figure~\ref{fig:background} and Table~\ref{tab:Background} 
show results of the fit with the background polynomial function 
to our sideband sample. 
In cross-checks with subsamples we find the variation of the shape parameters
is small compared to their statistical uncertainties.  
We use the nominal background shape with fixed parameters
in fits to data, and allow the parameters to vary constrained
by their errors as a systematic check.

\begin{table}[!htb]
\caption{\label{tab:Background} Fit parameters for the background shape from the fit to the sideband region.}
\begin{center}
\begin{tabular}{ l c }
\hline
\hline
Parameter             & Nominal value\\    
\hline
  $B_1$               &  0.63\PM0.22 \\
  $B_2$               &  0.95\PM0.39 \\
  $B_3$               &  0.41\PM0.54 \\
  $B_{xy}$            &--0.20\PM0.62 \\
  $B_{xyn}$           &--1.2\PM1.3   \\
  $B_{{\rm th,}x} \equiv B_{{\rm th,}y}$ 
                      &  1.31\PM0.13 \\
  $B_{{\rm th,}z}$    & 11.2\PM6.5   \\
\hline				   
Pearson $\chi^2/\nu$  & 129 / 97     \\
Probability (\%)      & 1.6          \\
Events on DP          & 1554         \\
\hline
\hline
\end{tabular}
\end{center}
\end{table}

\begin{verbatim}
\end{verbatim}

\section{Decay amplitude parametrization}
\label{sec:formalism}

\subsection{Matrix element}
\label{sec:matrix_element}

In this analysis we follow the formalism of E791 \cite{E791_Kpipi}
with only minor variations.
In the formulas below for the Dalitz plot variables, we also
use Mandelstam notations
$s=m^2(K^-\pi^+_1)$, 
$t=m^2(K^-\pi^+_2)$, and
$u=m^2(\pi^+_1\pi^+_2)$.
We choose two of them $s$ and $t$ as independent, and the third $u$ is dependent, constrained by the equation
$s+t+u=m_D^2 + m_K^2 + 2 m_\pi^2$. 
Then, the matrix element has an explicit Bose-symmetric form for pion permutations
\begin{equation}
\label{eqn:M}
  {\cal M}(s,t) = A(s,t) + A(t,s) + A_{L=0}^{I=2}(u(s,t)).
\end{equation}
Below we discuss the amplitudes contributing to the matrix element.

\subsection{\boldmath Partial $K\pi$ amplitudes}
\label{sec:partial_waves}

Each $K\pi$ amplitude is defined using a sum over the decay orbital momentum $L$
of two-body partial waves
\begin{equation}
\label{eqn:PWA_amplitude}
  A(s,t) = \sum_{L=0}^{L_{\rm max}} \Omega_L(s,t) 
                                    {\cal F}_D^L(q(s)) 
                                    {\cal A}_L(s),
\end{equation}
with parameters as described below. 
In this analysis we consider the sum up to 
the maximal orbital momentum $L_{\rm max}=2$.

\begin{figure}
    \includegraphics[width=80mm]{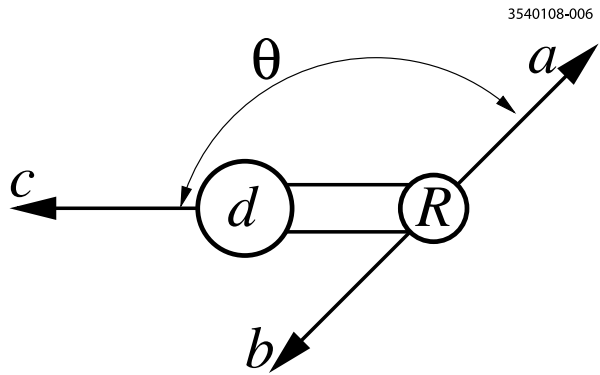}
    \caption{\label{fig:quasitwobody} Three-body decay $d \to Rc \to abc$
                                      in the resonance $R$ rest frame.}
\end{figure}
We assume the $D^+ \to K^-\pi^+\pi^+$ decay goes via a
quasi-two-body intermediate state,
$d \to R c$, containing the resonance $R$
and particle $c$, followed by the decay of the resonance
to the final stable particles $a$ and $b$, $R \to ab$.  This is
shown schematically in Fig.~\ref{fig:quasitwobody}.
The  $\Omega_L(s,t)$ term in Eq. (\ref{eqn:PWA_amplitude}) represents the angular distribution,
which we use in the invariant forms \cite{Tim}
\begin{eqnarray}
   \label{eqn:angular_distributions}
    \Omega_{L=0}(m^2_{ab},m^2_{ac})&=& 1, \label{eqn:angular_scalar} \\ 
    \Omega_{L=1}(m^2_{ab},m^2_{ac})&=& 
                 m^2_{bc} - m^2_{ac} + \frac{(m^2_d - m^2_c)(m^2_a - m^2_b)}{m^2_{ab}}, 
                                          \label{eqn:angular_vector}\\
    \Omega_{L=2}(m^2_{ab},m^2_{ac})&=&[\Omega_{L=1}]^2 
  - \frac{1}{3}\bigg( m^2_{ab} - 2 m^2_d - 2 m^2_c + \frac{(m^2_d - m^2_c)^2}{m^2_{ab}} \bigg)\nonumber \\
& & \times \bigg( m^2_{ab} - 2 m^2_a - 2 m^2_b + \frac{(m^2_a - m^2_b)^2}{m^2_{ab}} \bigg),
                                          \label{eqn:angular_tensor} 
\end{eqnarray}
where $m_d$, $m_a$, $m_b$, and $m_c$ are the masses of decaying and product
particles and $m_{ab}$,  $m_{ac}$, and $m_{bc}$
are the relevant invariant masses.
In the appendix 
we show that these angular distributions are equivalent 
to those applied in the E791 analysis \cite{E791_Kpipi} up to constant coefficients.

The form factors
${\cal  F}_D^L(q)$ in Eq.~(\ref{eqn:PWA_amplitude})
and ${\cal  F}_R^L(q)$ in Eqs.~(\ref{eqn:W_R}) and (\ref{eqn:Breit-Wigner_width})
are defined using the Blatt-Weisskopf form \cite{Blatt-Weisskopf}
\begin{eqnarray}
   \label{eqn:Blatt-W_formfactors}
   L=0: & & {\cal F}_{V}^0(q) = 1,                           \label{eqn:formf_scalar}\\
   L=1: & & {\cal F}_{V}^1(q) = \sqrt{\frac{1+q_V^2}{1+q^2}}, 
                                                               \label{eqn:formf_vector}\\
   L=2: & & {\cal F}_{V}^2(q) = \sqrt{\frac{9+3q_V^2+q_V^4}{9+3q^2+q^4}},  
                                                               \label{eqn:formf_tensor}
\end{eqnarray}
where the index $V$ stands for the $D$ or $R$ decay vertex,
$q=r_V {\rm P}$, $\rm P$ is the magnitude of the momentum of the 
decay products in the decaying particle's rest frame, and 
$r_V$ is the effective radius for the $D$ or $R$ vertex, respectively.  
For both $D$ and $R$ decays, $q_V = r_V{\rm P}_V$, where ${\rm P}_V$ is the magnitude of momentum of the decay products calculated at $m_{ab}=m_R$, 
the pole mass of $R$. 
The form factors are normalized by the condition 
${\cal F}_{V}^L(q_V) = 1$. 

The values of radial parameters are discussed in Sec.~\ref{sec:isobar_model}.
Expressions for the decay products' momentum for both vertices
can be found in the appendix. 
The mass dependences of ${\cal F}_D^L(q(s))$ form factors for
$K\pi$ resonances are shown in Table~\ref{tab:binning_scheme_Kpi}.

Another Gaussian form factor shape of the scalar resonance,
\begin{equation}
\label{eqn:Gaussian_form_factor}
 {\cal  F}_{V}^0(q) =  e^{-(q^2-q_V^2)/12},
\end{equation}
is applied in the E791~\cite{E791_Kpipi} analysis. 
This shape is suggested by Tornquist \cite{Tornquist}, and
has a steep dependence on momentum.
A reanalysis of the E791 data
\cite{Bugg-2006} found that this form factor is not required by the data.
We use only the Gaussian form factors ${\cal F}_{D}^0$ and ${\cal F}_{R}^0$ 
from Eq.~(\ref{eqn:Gaussian_form_factor}) 
for the scalar components $\kappa$ and $K^*_0(1430)$, 
when comparing results with E791 model~C \cite{E791-2002}.
We use the unit form factor from Eq.~(\ref{eqn:formf_scalar})
for scalar resonances in all other models and with the binned partial waves
discussed below.

The partial waves ${\cal A}_L(s)$ in Eq.~(\ref{eqn:PWA_amplitude})
are the angular momentum $L$-dependent functions of a single variable $z$, which is either $s$ or $t$. 
In the $D^+\to K^-\pi^+\pi^+$ decay
the $S$, $P$ and $D$ waves ($L=0$, 1, or 2, respectively)
are represented by the sum of functions $W_R$ for individual 
intermediate states 
\begin{eqnarray}
\label{eqn:A_0}
   {\cal A}_0(z) & = & c_{NR}                 
                     + W_{\kappa}
                     + W_{K_0^*(1430)}
                     + W_{S,{\rm binned}}, \\
\label{eqn:A_1}
   {\cal A}_1(z) & = & W_{K^*(892)}
                     + W_{K^*(1410)}
                     + W_{K^*(1680)}
                     + W_{P,{\rm binned}}, \\
\label{eqn:A_2}
   {\cal A}_2(z) & = & W_{K_2^*(1430)} 
                     + W_{D,{\rm binned}}. 
\end{eqnarray}
The contribution of nonresonant decays
is represented by $c_{NR}=a_{NR} e^{i\phi_{NR}}$, a complex constant
with two fit parameters for magnitude $a_{NR}$ and phase $\phi_{NR}$.
The $W_{L,{\rm binned}}$ are the binned amplitudes as discussed below.
\begin{equation}
\label{eqn:W_R}
   W_R = c_{R}   {\cal  W}_{R}   {\cal  F}_R^L(r_R {\rm P})
\end{equation}
is the shape of an individual resonance ${\cal  W}_{R}$ 
[see Eqs.~(\ref{eqn:Breit-Wigner})--(\ref{eqn:complex_pole})]
multiplied by the form factor in the resonance $R$ decay vertex
${\cal  F}_R^L(r_R {\rm P})$
and the coupling constant $c_R=a_R e^{i\phi_R}$.
The resonance $R$ production magnitudes $a_R$ and phases $\phi_R$ are parameters of
the fit to the Dalitz plot.

\subsection{Resonance shapes}
\label{sec:resonance_shapes}

For intermediate $K\pi$ resonances
we use the standard Breit-Wigner function
\begin{equation}
\label{eqn:Breit-Wigner}
{\cal  W}_R(m) = \frac{1}{m_R^2 - m^2 - i m_R \Gamma(m)},
\end{equation}
where $m^2=s$ 
and the mass-dependent width has the usual form
\begin{equation}
\label{eqn:Breit-Wigner_width}
 \Gamma(m) = \Gamma_R  
                       \frac{m_R}{m}
                       \bigg(\frac{\rm P}{{\rm P}_R} \bigg)^{2L+1} 
                       \big[{\cal F}_R^L(r_R {\rm P}) \big]^2.
\end{equation}

For 
$K_0^*(1430)$ we have tested both the Breit-Wigner function
[Eq.~(\ref{eqn:Breit-Wigner})] and the Flatt\'e parametrization
\begin{equation}
\label{eqn:Flatte}
{\cal  W}_R(m) = \frac{1}{m_R^2 - m^2 - i \sum_{ab} g^2_{Rab} \rho_{ab}(m)},
\end{equation}
where $g_{Rab}$ is a coupling constant of resonance $R$ to the final state $ab$ and
$\rho_{ab}(m) = 2{\rm P}/m$ is a phase space factor.
We test the Flatt\'e parametrization because
the $K_0^*(1430)$ mass is close to the $K\eta$ and $K\eta^\prime$  thresholds,
which could significantly distort the resonance shape~\cite{Bugg-2006}.

and the complex pole proposed in Ref.~\cite{Oller-2005}
\begin{equation}
\label{eqn:complex_pole}
{\cal  W}_R(m) = \frac{1}{s_R - m^2},
\end{equation}
where 
$s_R$ is a pole position in the complex $s=m^2$ plane.
This function represents the first term of the Laurent series
in the expansion of the chiral perturbation theory complex amplitude for the scalar wave.
This approach is common, and a survey of pole positions
extracted from different experiments can also be found in Ref.~\cite{Bugg-2005}.
This complex pole is equivalent to a Breit-Wigner function with constant width. 

\subsection{\boldmath Isospin-two $\pi^+\pi^+$ $S$ wave amplitude}
\label{sec:I2_pipi_wave}

The isospin-two $\pi^+\pi^+$ $S$ wave amplitude
in Eq.~(\ref{eqn:M}) is a sum of two components
\begin{equation}
\label{eqn:A_0_I2}
 A_{L=0}^{I=2}(u) =  c_{\pi\pi}{\cal W}_S^{I=2}
                   + W_{S,{\rm binned}}^{I=2},
\end{equation}
where $c_{\pi\pi}=a_{\pi\pi} e^{i\phi_{\pi\pi}}$ is a complex coupling constant and
$W_{S,{\rm binned}}^{I=2}$ is discussed in Sec.~\ref{sec:binned_amplitude}.
The first term of this sum is parametrized 
by a unitary form \cite{Achasov_PRD67_2003}
\begin{equation}
\label{eqn:I2pipiSwave}
     {\cal W}_S^{I=2}(m) = \frac{\eta_0^2(m)  e^{2 i \delta_0^2(m)} -1 }{2i},
\end{equation}
where $m$ is a $\pi^+\pi^+$ invariant mass,
$\eta_0^2(m)$ is an inelasticity, and $\delta_0^2(m)$ is a phase
of the $\pi^+\pi^+$ wave with total spin $0$ and isospin two. 
The phase $\delta_0^2(m)$ is assumed to be
proportional to the decay momentum at threshold and
sculpted by a polynomial function at higher mass range
\begin{equation}
   \label{eqn:delta_I2}
   \delta_0^2(m) = \frac{-a\sqrt{m^2/4-m_{\pi}^2}}{1 + bm^2 + cm^4 + dm^6}.
\end{equation}
with parameters 
  $a = (55.21 \pm 3.18)$~deg/GeV,
  $b = (0.853 \pm 0.254)$~GeV$^{-2}$,
  $c = (-0.959 \pm 0.247)$~GeV$^{-4}$, and
  $d = (0.314 \pm 0.070)$~GeV$^{-6}$,
obtained in Ref.~\cite{Achasov_PRD67_2003} from a fit to
the data of the scattering experiments \cite{Hoodland}.
We use this function with fixed parameters.
The inelasticity $\eta_0^2(m)$
in the mass range of $m < m_{\rm min} \approx m(\rho^+\rho^+) \approx $1.5~GeV/$c^2$
is expected to be near unity.
Then $\eta_0^2(m)$ decreases due to 
the $\pi^+\pi^+ \to \rho^+\rho^+$ rescattering at a higher mass range.
In our fits we use a smooth approximation for this threshold behavior
\begin{equation}
\label{eqn:eta02}
             \eta_0^2(m) = \left\{
\begin{array}{ll}
        1,                         & m \leq m_{\rm min}  \\
        1-\frac{\Delta\eta}{2}\bigg[1-\cos\bigg(\pi   \frac{m-m_{\rm min}}
                                                     {m_{\rm max}-m_{\rm min}}\bigg)\bigg], 
                                   & m_{\rm min}<m<m_{\rm max}                      \\
        1-\Delta\eta, 
                                   & m \geq m_{\rm max}, \\
\end{array} \right.
\end{equation}
with fit parameters $m_{\rm min}$, $m_{\rm max}$, and $\Delta\eta$.

\subsection{Binned amplitude}
\label{sec:binned_amplitude}

The complex term 
$W_{L,\rm binned}$ in Eqs.~(\ref{eqn:A_0})--(\ref{eqn:A_2}), and (\ref{eqn:A_0_I2}), 
where $L$=0, 1, or 2,
is intended to provide a completely model-free parametrization
of the partial wave. It can be used alone or in combination with
other terms. In the latter case it represents a correction to the complex 
amplitude of the isobar model.
We use this term in the form of an $s$-dependent complex number
\begin{equation}
\label{eqn:binned_wave}
 W_{L,\rm binned}(s) = a_L(s) e^{i\phi_L(s)},
\end{equation}
with functions $a_L(s)$ and $\phi_L(s)$ 
defined by an interpolation between the bins for the magnitude, $a_{Lk}$, 
and phase, $\phi_{Lk}$, where $k(s)=1,2,\ldots,N_L$ is 
an $s$-dependent index of these bins.
For all $K\pi$ waves we define $N_L=26$ uniform bins in $s \equiv m^2_{K\pi}$ 
in the range [0.4,3.0] (GeV/$c^2$)$^2$, as shown in 
Table~\ref{tab:binning_scheme_Kpi}.
Similar, for $I=2$ $\pi^+\pi^+$ $S$ wave we define $N_{L=0}^{I=2}=18$ uniform bins in 
$u \equiv m^2_{\pi\pi}$ [$s$ in Eq.~(\ref{eqn:binned_wave}) is replaced with $u$]
in the range [0.1,1.9]~(GeV/$c^2$)$^2$.
This binning scheme covers the kinematically allowed range 
of the $K\pi$ [0.633,1.730]~GeV/$c^2$ and $\pi^+\pi^+$ [0.279,1.376]~GeV/$c^2$
invariant mass spectrum in the $D^+ \to K^-\pi^+\pi^+$ decay.
We interpolate linearly between bin centers in our fitting function.

\begin{table}[!htb]
\caption{\label{tab:binning_scheme_Kpi} Bins for the $K\pi$ $S$, $P$, and $D$ waves and 
         Blatt-Weisskopf form factors for  
         $K^*(892)$ and $K^*(1680)$ from Eq.~(\ref{eqn:formf_vector}), 
         $K_2^*(1430)$ from Eq.~(\ref{eqn:formf_tensor}), and
         $K_0^*(1430)$ from Eq.~(\ref{eqn:Gaussian_form_factor}), 
         calculated in the $D$-meson decay vertex.}
\begin{center}
\begin{tabular}{ c c c c c c c c }
\hline
\hline
  \multicolumn{1}{c|}{Bin} 
& \multicolumn{1}{c|}{$m^2_{K\pi}$ (GeV/$c^2$)$^2$}  
& \multicolumn{2}{c|}{$m_{K\pi}$ (GeV/$c^2$)}                                    
& \multicolumn{3}{c|}{Blatt-W. form factors ${\cal F}_D^L(q)$ for }                
& Gaussian FF \\  
\cline{3-7}
\#  
& \multicolumn{1}{|c|}{Bin range}   
& \multicolumn{1}{c|}{Bin range}
& \multicolumn{1}{c|}{Center }
& \multicolumn{1}{c|}{$K^*(892)$ } 
& \multicolumn{1}{c|}{$K^*(1680)$} 
& \multicolumn{1}{c|}{$K_2^*(1430)$} 
& \multicolumn{1}{c}{$K_0^*(1430)$} \\
\hline
  1 & 0.4 --- 0.5 & 0.632 --- 0.707 & 0.671 & 0.888 & 0.250 & 0.305 & 0.347 \\
  2 & 0.5 --- 0.6 & 0.707 --- 0.775 & 0.742 & 0.918 & 0.259 & 0.324 & 0.380 \\
  3 & 0.6 --- 0.7 & 0.775 --- 0.837 & 0.806 & 0.948 & 0.267 & 0.345 & 0.415 \\
  4 & 0.7 --- 0.8 & 0.837 --- 0.894 & 0.866 & 0.982 & 0.277 & 0.368 & 0.451 \\
  5 & 0.8 --- 0.9 & 0.894 --- 0.949 & 0.922 & 1.017 & 0.287 & 0.394 & 0.489 \\
  6 & 0.9 --- 1.0 & 0.949 --- 1.000 & 0.975 & 1.055 & 0.297 & 0.421 & 0.528 \\
  7 & 1.0 --- 1.1 & 1.000 --- 1.049 & 1.025 & 1.096 & 0.309 & 0.452 & 0.570 \\
  8 & 1.1 --- 1.2 & 1.049 --- 1.095 & 1.072 & 1.140 & 0.321 & 0.485 & 0.612 \\
  9 & 1.2 --- 1.3 & 1.095 --- 1.140 & 1.118 & 1.188 & 0.335 & 0.523 & 0.656 \\
 10 & 1.3 --- 1.4 & 1.140 --- 1.183 & 1.162 & 1.240 & 0.349 & 0.564 & 0.700 \\
 11 & 1.4 --- 1.5 & 1.183 --- 1.225 & 1.204 & 1.296 & 0.365 & 0.609 & 0.746 \\
 12 & 1.5 --- 1.6 & 1.225 --- 1.265 & 1.245 & 1.358 & 0.383 & 0.659 & 0.792 \\
 13 & 1.6 --- 1.7 & 1.265 --- 1.304 & 1.285 & 1.425 & 0.402 & 0.715 & 0.839 \\
 14 & 1.7 --- 1.8 & 1.304 --- 1.342 & 1.323 & 1.499 & 0.423 & 0.776 & 0.885 \\
 15 & 1.8 --- 1.9 & 1.342 --- 1.378 & 1.360 & 1.581 & 0.446 & 0.844 & 0.932 \\
 16 & 1.9 --- 2.0 & 1.378 --- 1.414 & 1.396 & 1.672 & 0.471 & 0.918 & 0.978 \\
 17 & 2.0 --- 2.1 & 1.414 --- 1.449 & 1.432 & 1.773 & 0.500 & 0.999 & 1.023 \\
 18 & 2.1 --- 2.2 & 1.449 --- 1.483 & 1.466 & 1.886 & 0.532 & 1.085 & 1.067 \\
 19 & 2.2 --- 2.3 & 1.483 --- 1.517 & 1.500 & 2.013 & 0.568 & 1.178 & 1.109 \\
 20 & 2.3 --- 2.4 & 1.517 --- 1.549 & 1.533 & 2.157 & 0.608 & 1.275 & 1.150 \\
 21 & 2.4 --- 2.5 & 1.549 --- 1.581 & 1.565 & 2.320 & 0.654 & 1.374 & 1.189 \\
 22 & 2.5 --- 2.6 & 1.581 --- 1.612 & 1.597 & 2.506 & 0.707 & 1.472 & 1.226 \\
 23 & 2.6 --- 2.7 & 1.612 --- 1.643 & 1.628 & 2.719 & 0.766 & 1.568 & 1.259 \\
 24 & 2.7 --- 2.8 & 1.643 --- 1.673 & 1.658 & 2.962 & 0.835 & 1.657 & 1.290 \\
 25 & 2.8 --- 2.9 & 1.673 --- 1.703 & 1.688 & 3.240 & 0.913 & 1.737 & 1.318 \\
 26 & 2.9 --- 3.0 & 1.703 --- 1.732 & 1.718 & 3.554 & 1.002 & 1.805 & 1.342 \\
\hline
\multicolumn{4}{c}{$m_{K\pi,~\rm min}=m_{K^-} + m_{\pi^+}$ = 0.633~GeV/$c^2$} & 
\multicolumn{4}{c}{}\\ 
\multicolumn{4}{c}{$m_{K\pi,~\rm max}=m_{D^+} - m_{\pi^+}$ = 1.730~GeV/$c^2$} & 
\multicolumn{4}{c}{}\\
\hline
\hline
\end{tabular}
\end{center}
\end{table}

\subsection{Fit fraction} 
\label{sec:fit_fraction}

We estimate a contribution of each component in the matrix element
using a standard definition of the fit fraction,
\begin{equation}
\label{eqn:FFR}
  {\rm FF}_R = \frac{\int  |A_R(x,y)|^2 dx dy}{\int  |M(x,y)|^2 dx dy}, 
\end{equation}
where $A_R(x,y)$ is an amplitude contribution from the $R$ component 
to the total matrix element $M(x,y)$ from Eq.~(\ref{eqn:M}) and the integrals
are taken over the area of the Dalitz plot.

\subsection{Expected contributions} 
\label{sec:expected_contributions}

\begin{figure}[!htb]
  \includegraphics[width=160mm]{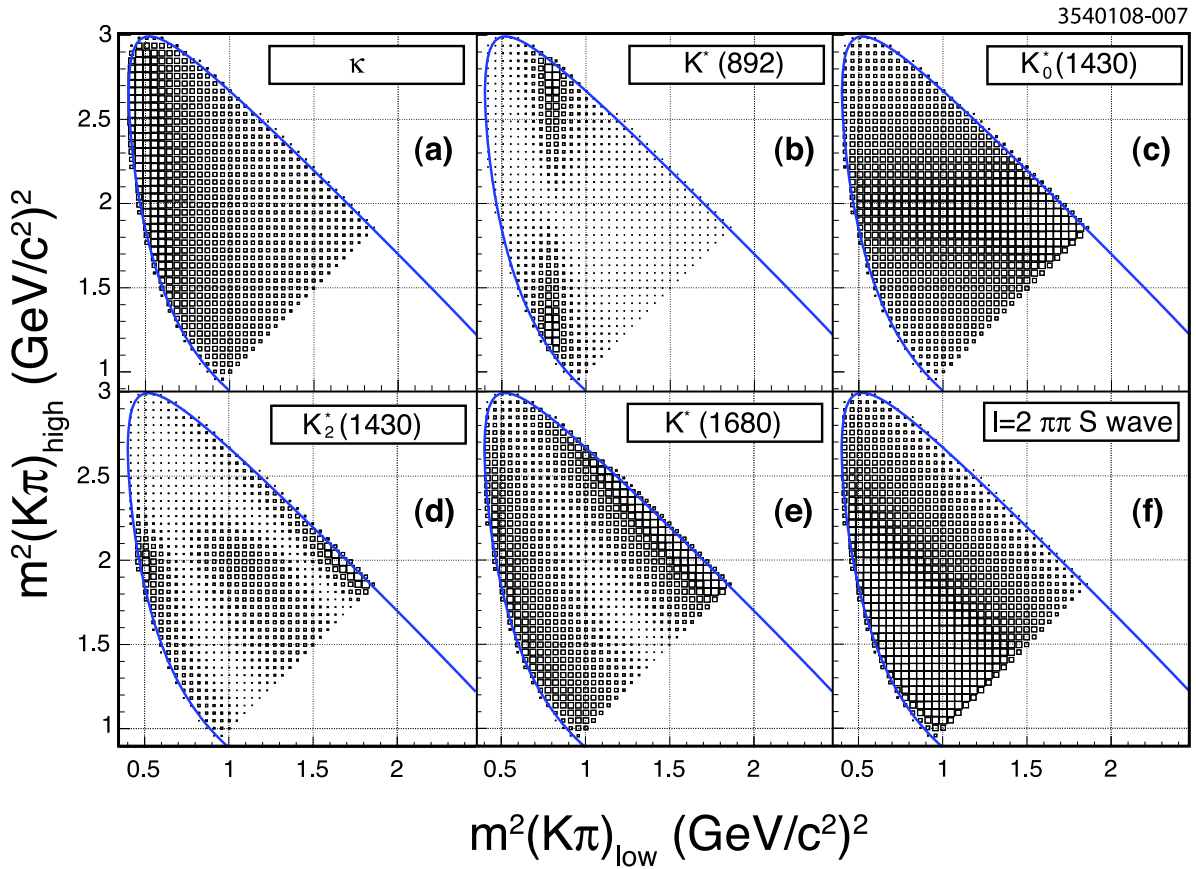}
  \caption{\label{fig:expected} Simulation of the expected contribution to the $D^+ \to K^-\pi^+\pi^+$
                                Dalitz plot from various intermediate states.
                                (a) Low mass $K\pi$ $S$ wave ($\kappa$),
                                (b) $K^*(892)\pi$,         
                                (c) $K^*_0(1430)\pi$,         
                                (d) $K^*_2(1430)\pi$,         
                                (e) $K^*(1680)\pi$, and 
                                (f) $K(\pi\pi)_{I=2}$ with $I=2$ $\pi^+\pi^+$ $S$ wave.          
                                } 
\end{figure}

{\it A priori}, all known $K\pi$ resonances in the mass range 
from the production threshold up to 1.73~GeV/$c^2$,
such as
$K^*(892)$,
$K^*(1410)$,
$K_0^*(1430)$,
$K_2^*(1430)$,
$K^*(1680)$, and even higher mass resonances,
might contribute to the amplitude of the $D^+\to K^-\pi^+\pi^+$ decay.
Table~\ref{tab:model_parameters} shows 
parameters of $K\pi$ resonances
which have been considered in this analysis.
One would also foresee an $I=2$ $\pi^+\pi^+$ $S$ wave final state interaction.
Simulations of some of the expected contributions to the Dalitz plot are shown in
Fig.~\ref{fig:expected}. In contrast to data
[Fig.~\ref{fig:DP_data}(a)] 
the $K^*(892)$ population density in Fig.~\ref{fig:expected}(b) is symmetric
without interference with an $S$ wave.

\section{Fits to data using the isobar model}
\label{sec:isobar_model}

\subsection{Comparison with E791 Model~C}
\label{sec:comparison_with_E791}

\begin{figure}[!htb]
  \includegraphics[width=160mm]{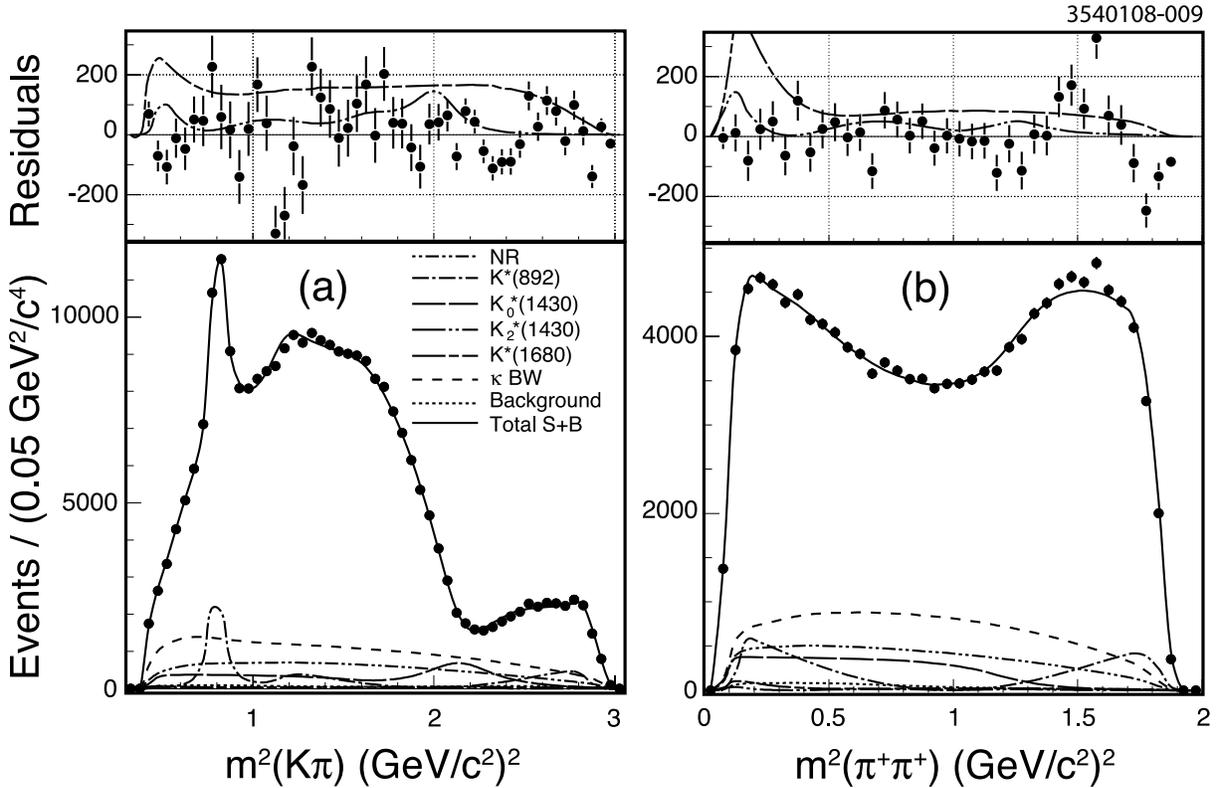}
  \caption{\label{fig:comp_E791_Model_C} Projections of the fit to the Dalitz plot with model~C on
                                         (a) $m^2(K\pi)$ (two entries per event) and 
                                         (b) $m^2(\pi\pi)$ variables. Residuals between data and
                                         the total p.d.f.\ are shown by dots 
                                         with statistical error bars on the
                                         top insets along with minor contributions from the 
                                         $K^*(1680)$ and $K_2^*(1430)$ resonances, plotted with factor 
                                         $\times$4.} 
\end{figure}

First we compare our results, obtained in the framework of the isobar model,
with E791 models A, B, and C from Ref.~\cite{E791-2002}.
In particular, the most complete model C contains
$\bar K^*(892)\pi^+$,
$\bar K_0^*(1430)\pi^+$,
$\bar K_2^*(1430)\pi^+$,
$\bar K^*(1680)\pi^+$,
$\kappa\pi^+$, and nonresonant (NR) contributions.
Following E791 we allow a scalar $K\pi$ amplitude,
the ``$\kappa$,'' as a Breit-Wigner resonance with mass-dependent width.
We set $c_{K^*(892)}=1$ in Eqs.~(\ref{eqn:A_1}) and (\ref{eqn:W_R}),
and all other magnitudes and phases 
are defined with respect to $K^*(892)$.
Gaussian form factors ${\cal F}_D^0(r_D {\rm P}_c)$ and ${\cal F}_R^0(r_R {\rm P})$  
from Eq.~(\ref{eqn:Gaussian_form_factor})
are used for $K_0^*(1430)$ and $\kappa$.
For all $K\pi$ resonances with nonzero spin,
the radii in the Blatt-Weisskopf~\cite{Blatt-Weisskopf} form factors 
$r_D=5$~GeV$^{-1}$ and $r_R=1.5$~GeV$^{-1}$ are fixed to the values
used by E791.  
On the Dalitz plot, the p.d.f.\ for this model  
looks indistinguishable from statistics
shown in Fig.~\ref{fig:DP_data}(a). 
The Dalitz plot projections with p.d.f.\ components are shown in 
Fig.~\ref{fig:comp_E791_Model_C}.
Fit fractions and phases, obtained in our fit, are compared with E791~\cite{E791-2002}
in Table~\ref{tab:modelC} and are statistically consistent.
Magnitudes are not comparable because of a different choice for normalization.
The values obtained for resonance parameters are compared in Table~\ref{tab:model_parameters}.
In particular, we get dominant contributions from $S$ wave components:
the $NR$, $K_0^*(1430)\pi$, and $\kappa\pi$ fit fractions are 
$\approx$9\%, 10\%, and 33\%, respectively. 
The total sum of all fit fractions
is 65.5\%, indicating substantial constructive interference.
Apparently, model~C gives a poor fit quality $\chi^2/\nu$=531/391.
A large discrepancy between the fit and the data is seen in
Fig.~\ref{fig:comp_E791_Model_C} for the $m^2(\pi^+\pi^+)$ projection
in the range of [1.4,~1.9] (GeV/$c^2$)$^2$. 
That motivates us to explore alternative models of the decay amplitude.

\begin{table}[!htb]
\caption{\label{tab:model_parameters} 
         The Breit-Wigner resonance parameters used or measured in the isobar model;
         CLEO-c vs E791. The fixed parameters used in the E791 experiment are taken
         from PDG 2000~\cite{PDG-2000} with their uncertainties shown in square brackets.
         The measured values are shown with two uncertainties: statistical and systematic.
         The values shown in parentheses (with statistical error only) 
         were obtained in cross-checks when these parameters are allowed to float. 
         }         
\begin{center}
\begin{tabular}{ c c c c c }
\hline
\hline
  Parameter   
& \multicolumn{1}{|c|}{E791 [PDG 2000] }
& \multicolumn{2}{ c|}{CLEO-c}    
& PDG 2006 \cite{PDG-2006} \\  
\cline{2-4}
  (MeV/$c^2$) 
& \multicolumn{2}{|c|}{ Model C (if float)  } 
& \multicolumn{1}{ c|}{ Model~I2 (if float) } & \\
\hline
$m_{K^*(892)}$                     & 896.1 [\PM0.27] 
                                   & 896   (894.8\PM0.5) 
                                   & 895.7\PM0.2\PM0.3          
                                   & 896.00\PM0.25  \\
$\Gamma_{K^*(892)}$                & 50.7 [\PM0.6] 
                                   & 50.3 (45.5\PM0.4)   
                                   & 45.3\PM0.5\PM0.6        
                                   & 50.3\PM0.6     \\
$m_{K^*(1430)}$                    & 1459\PM7\PM12  
                                   & 1463.0\PM0.7\PM2.4       
                                   & 1466.6\PM0.7\PM3.4       
                                   & 1414\PM6       \\ 
$\Gamma_{K^*(1430)}$               & 175\PM12\PM12
                                   & 163.8\PM2.7\PM3.1      
                                   & 174.2\PM1.9\PM3.2       
                                   & 290\PM21       \\
$m_{K_2^*(1430)}$                  & 1432.4 [\PM1.3]   
                                   & 1432.4 (1436\PM11)   
                                   & 1432.4 (1427\PM7)      
                                   & 1432.4\PM1.3   \\ 
$\Gamma_{K_2^*(1430)}$             & 109 [\PM5]   
                                   & 109 (132\PM21) 
                                   & 109 (120\PM13)     
                                   & 109\PM5        \\
$m_{K^*(1680)}$                    & 1717 [\PM27]   
                                   & 1717 (1782\PM41)     
                                   & 1717 (1679\PM59) 
                                   & 1717\PM27      \\ 
$\Gamma_{K^*(1680)}$               & 322 [\PM110]  
                                   & 322 (565\PM131)     
                                   & 322 (446\PM119)
                                   & 322\PM110      \\
$m_{K^*(1410)}$                    & 1414[\PM15]
                                   & 1414    
                                   & 1414    
                                   & 1414\PM15      \\ 
$\Gamma_{K^*(1410)}$               & 232[\PM21]
                                   & 232   
                                   & 232   
                                   & 232\PM21       \\
$m_{\kappa}$                       & 797\PM19\PM43
                                   & 809\PM1\PM13        
                                   & Complex pole,
                                   & $K_0^*(800)$ is not  \\
$\Gamma_\kappa$                    & 410\PM43\PM87   
                                   & 470\PM9\PM15        
                                   & see Table~\ref{tab:alternative_parameters}
                                   & established    \\
\hline
\hline
\end{tabular}
\end{center}
\end{table}

\begin{table}[!htb]
\caption{\label{tab:modelC} 
Comparison of CLEO-c results with E791 using the isobar fit, model~C. 
Shown are the fitted magnitudes, $a$ in arbitrary units, the phases, $\phi$ in degrees, 
defined relative to the $K^*(892)\pi^+$ amplitude, and the FF. 
Also indicated are the fitted masses $m$ and widths $\Gamma$ of the 
spin-zero resonances.
Magnitudes $a$ are not comparable between the two experiments because of a different choice for normalization.
}
\begin{center}
\begin{tabular}{ l l c c}
\hline
\hline
Mode & Parameter        & E791               & CLEO-c                \\  
\hline     
NR   & $a$              & 1.03\PM0.30\PM0.16 & 7.4\PM0.1\PM0.6       \\ 
     & $\phi$ $(^\circ)$& --11\PM14\PM8      & --18.4\PM0.5\PM8.0    \\  
     & FF   (\%)        & 13.0\PM5.8\PM4.4   & 8.9\PM0.3\PM1.4       \\ 
\hline							                
${\overline K}^*(892)\pi^+$				                
     & $a$              & 1 (fixed)          & 1 (fixed)            \\ %
     & $\phi$ $(^\circ)$& 0 (fixed)          & 0 (fixed)            \\ %
     & FF   (\%)        & 12.3\PM1.0\PM0.9   & 11.2\PM0.2\PM2.0     \\ %
\hline							                
${\overline K}_0^*(1430)\pi^+$   			                
     & $a$              & 1.01\PM0.10\PM0.08 & 3.00\PM0.06\PM0.14   \\ %
     & $\phi$ $(^\circ)$& 48\PM7\PM10        & 49.7\PM0.5\PM2.9     \\ %
     & FF   (\%)        & 12.5\PM1.4\PM0.5   & 10.4\PM0.6\PM0.5     \\ %
     & m (MeV/$c^2$)    & 1459\PM7\PM12      & 1463.0\PM0.7\PM2.4   \\ %
     &$\Gamma$ (MeV/$c^2$)&175\PM12\PM12     & 163.8\PM2.7\PM3.1    \\ %
\hline							                
${\overline K}_2^*(1430)\pi^+$   			                
     & $a$              & 0.20\PM0.05\PM0.04 & 0.962\PM0.026\PM0.050\\ %
     & $\phi$ $(^\circ)$& --54\PM8\PM7       & --29.9\PM2.5\PM2.8   \\ %
     & FF   (\%)        & 0.5\PM0.1\PM0.2    & 0.38\PM0.02\PM0.03   \\ %
\hline							                
${\overline K}^*(1680)\pi^+$   				                
     & $a$              & 0.45\PM0.16\PM0.02 & 6.5\PM0.1\PM1.5      \\ %
     & $\phi$ $(^\circ)$& 28\PM13\PM15       & 29.0\PM0.7\PM4.6     \\ %
     & FF   (\%)        & 2.5\PM0.7\PM0.3    & 1.28\PM0.04\PM0.28   \\ %
\hline							                
$\kappa \pi^+$   				                        
     & $a$              & 1.97\PM0.35\PM0.11 & 5.01\PM0.04\PM0.27   \\ %
     & $\phi$ $(^\circ)$& --173\PM8\PM18     & --163.7\PM0.4\PM5.8  \\ %
     & FF   (\%)        & 47.8\PM12.1\PM5.3  & 33.2\PM0.4\PM2.4     \\ %
     & m (MeV/$c^2$)    & 797\PM19\PM43      & 809\PM1\PM13         \\ %
     &$\Gamma$ (MeV/$c^2$)&410\PM43\PM87     & 470\PM9\PM15         \\ %
\hline							                
Form factor						                
     & $r_{\kappa}($GeV$^{-1})$&1.6\PM1.3    & 1.5(fixed)           \\ %
     & $r_D($GeV$^{-1})$& 5.0\PM0.5          & 5  (fixed)           \\ %
Other $R \to K\pi$					                
     & $r_R($GeV$^{-1})$& 1.5(fixed)         & 1.5(fixed)           \\ %
\hline							                
$\sum$ FF (\%) &        & 88.6               & 65.5                 \\ 
Goodness						                
     & $\chi^2 / \nu$   & 46/63              & 531/391              \\ %
\hline
\hline
\end{tabular}
\end{center}
\end{table}

\begin{table}[!htb]
\caption{\label{tab:alternative_parameters} Alternative parameters 
                         obtained in the fits with the isobar model.}
\begin{center}
\begin{tabular}{l l l c c }
\hline
\hline
Mode & 
Amplitude & Parameter (MeV/$c^2$)   & Model~C            & Model~I2        \\ 
\hline					                       
${\overline K}_0^*(1430)\pi^+$ 
& Breit-Wigner                                              
     & $m_{K_0^*(1430)}$       & 1463.0\PM0.7\PM2.4 & 1466.6\PM0.7\PM3.4   \\
&    & $\Gamma_{K_0^*(1430)}$  &  163.8\PM2.7\PM3.1 &  174.2\PM1.9\PM3.2   \\
\cline{2-5}
& Flatt\'e	
      & $m_{K_0^*(1430)}$       & 1462.5\PM3.9       & 1471.2\PM0.8        \\
&     & $g_{K\pi}$              & 532.9\PM8.5        & 546.8\PM4.2         \\
&     & $g_{K\eta}$             & 0                  & 0                   \\
&     & $g_{K\eta'}$            & 197\PM106          & 230\PM32            \\
\hline					               
$\kappa \pi^+$ 
& Breit-Wigner 
      & $m_{\kappa}$            & 809\PM1\PM13       & 888\PM2             \\
&     & $\Gamma_{\kappa}$       & 470\PM9\PM15       & 550\PM12            \\
\cline{2-5}
& Complex pole                                     
     & $\Re m_{\kappa}$         &  769.9\PM6.3       & 706.0\PM1.8\PM22.8  \\
&    & $\Im m_{\kappa}$         &--221.2\PM8.4       &--319.4\PM2.2\PM20.2 \\
\hline
\hline
\end{tabular}
\end{center}
\end{table}

\subsection{Variations of Model~C}
\label{sec:variations_of_modelC}

The Gaussian form factors, 
given by Eq.~(\ref{eqn:Gaussian_form_factor}) for scalar resonances,
${\cal F}_R^0(r_R {\rm P})$, and 
${\cal F}_D^0(r_D {\rm P}_c)$,
behave similarly to the $L=4$ Blatt-Weisskopf form factor. This behavior is 
not preferred by either our or the E791 data; see Ref.~\cite{Bugg-2006} for details. 
With our data
we find that results are not very dependent on the assumed resonance 
decay vertex form factor ${\cal F}_R^0(r_R {\rm P})$.
However, the $D$-meson decay vertex form factor
${\cal F}_D^0(r_D {\rm P})$ changes the $S$ wave dependence on $s$ significantly.
In particular, this Gaussian form factor (see Table~\ref{tab:binning_scheme_Kpi})
suppresses the contribution of $K\pi$ at low mass. 
To agree with the data when fitting with this factorized form factor, 
the magnitude of the complex function ($S$ wave)
increases at low $K\pi$ mass which gives an
illusion of resonance behavior.
For all models other than model~C
below we use unit form factors from Eq.~(\ref{eqn:formf_scalar}) for $S$ wave contributions.

For the $K_0^*(1430)$ resonance we measure
$m_{K_0^*(1430)}=1463.0 \pm 0.7 \pm 2.4$~MeV/$c^2$ and 
$\Gamma_{K_0^*(1430)}=163.8 \pm 2.7 \pm 3.1$~MeV/$c^2$, 
which are consistent with E791 results,
but inconsistent with current PDG~\cite{PDG-2006} values,
as demonstrated in Table~\ref{tab:model_parameters}.
Similar behavior is reported in Ref.~\cite{FOCUS-2007}
from the FOCUS Collaboration.
In Ref.~\cite{Bugg-2006} Bugg surmises that the $K_0^*(1430)$ resonance parameters
might change due to the opening of the $K\eta'$ channel. 
In order to accommodate this effect, we test
the Flatt\'e parametrization of Eq.~(\ref{eqn:Flatte}), 
which depends on a floating mass $m_{K_0^*(1430)}$ and three coupling constants,
$g_{K\pi}$, $g_{K\eta}$, and $g_{K\eta'}$.  We find that our data are
consistent with $g_{K\eta} = 0$, and this coupling is dropped from further consideration. 
The resulting values of the other parameters are shown in 
Table~\ref{tab:alternative_parameters}. 
We do not find any significant difference between 
Breit-Wigner and Flatt\'e parametrizations 
in the shape of the $K_0^*(1430)$ complex amplitude or in the fit quality. 

In model~C we also measure the $\kappa$ resonance Breit-Wigner parameters
$m_{\kappa}=809 \pm 1\pm 13$~MeV/$c^2$ and
$\Gamma_{\kappa}=470 \pm 9\pm 15$~MeV/$c^2$.
In Ref.~\cite{Oller-2005} Oller emphasizes that the Breit-Wigner resonance 
with a mass-dependent width is not the 
best choice for the $K\pi$ low-mass phase parametrization.
Following his prescription, we replace the Breit-Wigner function by
the complex pole from Eq.~(\ref{eqn:complex_pole}) with the initial pole position at
$s_{\kappa} =(710 -i 310)^2$~MeV$^2/c^4$. We allow the pole position to float
and obtain its optimal location, as shown in Table~\ref{tab:alternative_parameters}.

We also test the $K^*(892)$ parameters,
as shown in Table~\ref{tab:model_parameters}. 
The $K^*(892)$ mass is consistent with the PDG~\cite{PDG-2006} value, 
while the width is about 5~MeV/$c^2$ smaller. 
In further analysis we allow the $K^*(892)$ mass and width to float.

Inclusion of a $K^*(1410)$ component in the fit does not result in 
       any significant contribution.

\subsection{Model I2}
\label{sec:model_I2}

Tuning of the isobar model for $K\pi$ waves only does not improve 
significantly the probability of consistency between the data and 
the model, which is still small. 
The large discrepancy in the $m^2(\pi^+\pi^+)$ spectrum 
persists even if we use the model-independent parametrization 
for the $K\pi$ $S$ wave considered below.
We solve this problem by 
including in the matrix element [Eq.~(\ref{eqn:M})] 
a contribution from the $I=2$ $\pi^+\pi^+$ $S$ wave,
using Eq.~(\ref{eqn:I2pipiSwave}), which leads to a model called model~I2. 
The threshold of the $\pi^+\pi^+ \to \rho^+\rho^+$ rescattering process 
is located at the upper edge of the $m^2(\pi^+\pi^+)$ 
kinematic limit in our decay.  The edge of the kinematic border
does not allow us to resolve correlations between the
$\Delta\eta$ and $m_{\rm max}$ parameters in Eq.~(\ref{eqn:eta02}).
We fix the value of $\Delta\eta =1$, while
floating the $m_{\rm min}$ and $m_{\rm max}$ parameters. 
In model~I2, compared to model C, 
the unit form factors ${\cal F}_{D}^0 = {\cal F}_{R}^0 = 1$
are used for the scalar components, the complex pole amplitude 
from Eq.~(\ref{eqn:complex_pole}) is used
for the low-mass $K\pi$ $S$ wave, and the $K^*(892)$ parameters are allowed
to vary in the fit; other conditions stay the same as in model C. 

\begin{figure}[!htb]
  \includegraphics[width=160mm]{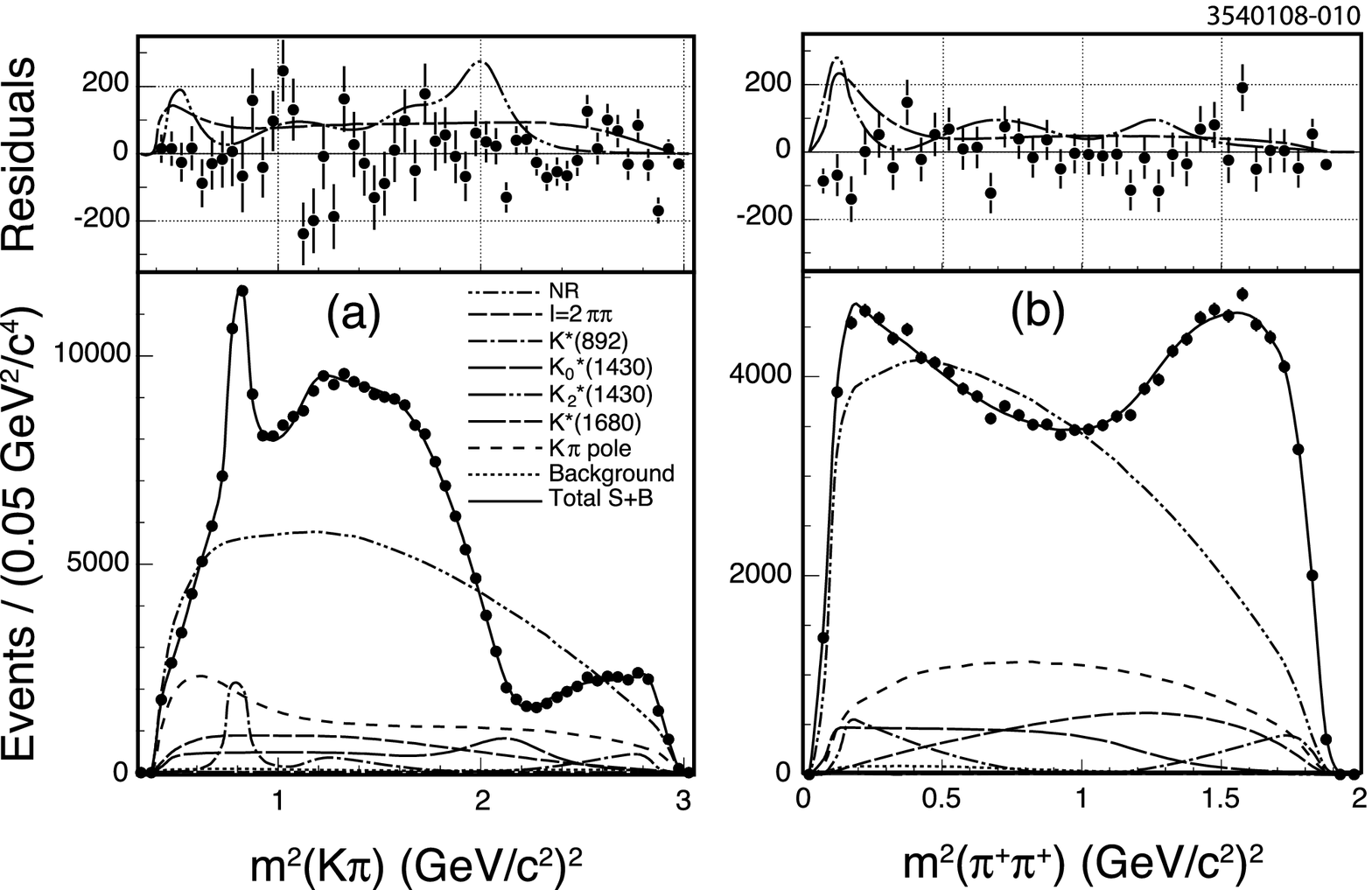}
  \caption{\label{fig:Projections_Model_I2} Projections of the fit to the Dalitz plot with model~I2 on 
                                         (a) $m^2(K\pi)$ (two entries per event) and 
                                         (b) $m^2(\pi\pi)$ variables. Residuals between data and
                                         the total p.d.f.\ are shown by dots 
                                         with statistical error bars on the
                                         top insets along with minor contributions from the 
                                         $K^*(1680)$ and $K_2^*(1430)$ resonances, plotted with factor 
                                         $\times$10.} 
\end{figure}

Model~I2 gives the best analytical description of our data.
The obtained fit parameters 
are presented in Table~\ref{tab:Isobar_I2_vs_QMIPWA_I2}.
The total p.d.f.\ and separate components 
are shown in the Dalitz plot projections in 
Fig.~\ref{fig:Projections_Model_I2}. 
The large discrepancy in the $m^2(\pi^+\pi^+)$ spectrum is eliminated, 
improving the probability of consistency between the model and data to 13\%.

\subsection{Variations of Model~I2}
\label{sec:variations_of_modelI2}

We did not find any significant contribution from 
$K^*(1410)$ by including it in the fit. 
This resonance is excluded from further consideration. 

We also test the alternative descriptions for the $K_0^*(1430)$
parameters as shown Table~\ref{tab:alternative_parameters}.
We do not find any significant difference between 
Breit-Wigner and Flatt\'e parametrizations 
in the shape of the $K_0^*(1430)$ complex amplitude or in the fit quality. 
We therefore use the Breit-Wigner function in model~I2
with the mass and width of $K_0^*(1430)$ allowed to float.

The test of alternative descriptions for the low-mass $K\pi$ $S$ wave 
is presented in Tables~\ref{tab:alternative_parameters} and
                       \ref{tab:Isobar_I2_vs_QMIPWA_I2}.
The pole amplitude slightly improves the fit quality ($\Delta\chi^2=-10$)
compared to the Breit-Wigner resonance with mass-dependent width. 
This substitution leads to a significant redistribution of 
the $NR$ and $\kappa$ components of the scalar wave, though their sum
results in very small variation of the complex function.
In model~I2 we use the pole amplitude for the $K\pi$ low-mass $S$ wave.

\begin{table}[!htb]
\caption{\label{tab:Isobar_I2_vs_QMIPWA_I2} Fit results for model~I2, with a Breit-Wigner
         function or with a complex pole for the $\kappa$, and QMIPWA.
         The FF are shown for a single $K\pi$ wave and need to be doubled
         as indicated by the ``$2 \times$'' symbol in row titles. }
\small
\begin{center}
\begin{tabular}{l l c c c}
\hline
\hline
Mode & Parameter               & Model~I2 (B-W for $\kappa$)
                                                    & Model~I2           & QMIPWA        \\  
\hline			                            
${\overline K}^*(892)\pi^+$                         		                    
     & $a$                     & 1 -- fixed         & 1 -- fixed         & 1 -- fixed    \\
     & $\phi$ $(^\circ)$       & 0 -- fixed         & 0 -- fixed         & 0 -- fixed    \\ 
     & FF   (\%) $2 \times$    & 5.15\PM0.24        & 5.27\PM0.08\PM0.15 & 4.94\PM0.23   \\  
     & $m$ (MeV/$c^2$)         & 895.4\PM0.2        & 895.7\PM0.2\PM0.3  & 895.7 -- fixed\\
     & $\Gamma$ (MeV/c$^2$)    &  44.5\PM0.7        &  45.3\PM0.5\PM0.6  &  45.3 -- fixed\\
\hline			                            		                    
${\overline K}^*(1680)\pi^+$                        		                    
     & $a$                     & 4.45\PM0.23        & 3.38\PM0.16\PM0.78 & 2.88\PM0.84   \\
     & $\phi$ $(^\circ)$       & 43.3\PM3.6         & 68.2\PM1.6\PM13    & 113\PM14      \\ 
     & FF   (\%) $2 \times$    &0.238\PM0.024       &0.144\PM0.013\PM0.12&0.098\PM0.059  \\  
\hline			                            		                    
${\overline K}_2^*(1430)\pi^+$                      		                    
     & $a$                     &0.866\PM0.030       &0.915\PM0.025\PM0.04& 0.794\PM0.073 \\
     & $\phi$ $(^\circ)$       &--17.4\PM3.5        &--17.4\PM2.3\PM2.0  & 14.8\PM9.0    \\ 
     & FF   (\%) $2 \times$    &0.124\PM0.011       &0.145\PM0.009\PM0.03& 0.102\PM0.020 \\  
\hline			                            		                    
${\overline K}_0^*(1430)\pi^+$                      		                    
     & $a$                     & 3.97\PM0.15        & 3.74\PM0.02\PM0.06 & 3.74 -- fixed \\
     & $\phi$ $(^\circ)$       & 45.1\PM0.9         & 51.1\PM0.3\PM1.6   & 51.1 -- fixed \\ 
     & FF   (\%) $2 \times$    & 7.53\PM0.65        & 7.05\PM0.14\PM0.55 & 6.65\PM0.31   \\  
     & $m$ (MeV/$c^2$)         & 1461.1\PM1.0       & 1466.6\PM0.7\PM3.4 &1466.6 -- fixed\\
     & $\Gamma$ (MeV/$c^2$)    &  177.9\PM3.1       &  174.2\PM1.9\PM3.2 & 174.2 -- fixed\\
\hline			                            		                    
$\kappa \pi^+$		                            
     & $a$                     & 5.69\PM0.17        & 10.80\PM0.05\PM0.35& 0	         \\
     & $\phi$ $(^\circ)$       &--149.9\PM1.2       &  148.4\PM0.3\PM1.6 & 0             \\ 
     & FF   (\%) $2 \times$    & 8.5\PM0.5          &  21.6\PM0.3\PM3.2  & 0		 \\  
 Pole   	       	                                         
     & $\Re m_0$ (MeV/$c^2$)   &		    &  706.0\PM1.8\PM22.8&               \\
     & $\Im m_0$ (MeV/$c^2$)   &		    &--319.4\PM2.2\PM20.2&               \\
 Breit-Wigner   	       	                                         
     & $m$ (MeV/$c^2$)         &  888.0\PM1.9       &		         &               \\
     & $\Gamma$ (MeV/$c^2$)    &  550.4\PM11.8      &		         &               \\
\hline			                            			    
NR   & $a$                     & 17.1\PM0.4         & 23.3\PM0.1\PM1.6   & 0             \\
     & $\phi$ $(^\circ)$       & 1.9\PM1.7          & 29.7\PM0.2\PM3.0   & 0             \\ 
     & FF   (\%)               & 38.0\PM1.9         & 73.8\PM0.8\PM9.6   & 0             \\  
\hline			                            		                    
Binned $K^- \pi^+$ $S$ wave                           	       	                         
     & $a$                     & 0                  & 0		         & See           \\
     & $\phi$ $(^\circ)$       & 0                  & 0		         & 
        Table~\ref{tab:results_Swave_with_I2}  \\ 
     & FF   (\%) $2 \times$    & 0                  & 0		         & 41.9\PM1.9    \\  
\hline			                            			    
$I=2$ $\pi^+\pi^+$ $S$ wave	                            		            
     & $a$                     & 30.3\PM2.7         & 25.5\PM0.3\PM2.9   & 33.1\PM2.6    \\
     & $\phi$ $(^\circ)$       & 86.3\PM3.3         & 75.4\PM0.6\PM10    & 66.2\PM3.5    \\ 
     & FF   (\%)               & 13.4\PM2.3         &  9.8\PM0.2\PM2.0   & 15.5\PM2.8    \\  
Equation~(\ref{eqn:eta02})	                            
     & $\Delta\eta$            & 1                  & 1                  & 1             \\
     &$m_{\rm min}$ (MeV/$c^2$)& 1265\PM8           & 1256\PM5\PM4       &1256.3 -- fixed\\
     &$m_{\rm max}$ (MeV/$c^2$)& 1529\PM31          & 1498\PM13\PM30     &1498.2 -- fixed\\
\hline			                            			    
Form factor		                            			    
     & $r_D($GeV$^{-1})$       & 5 -- fixed         & 5 -- fixed         & 5 -- fixed    \\
     & $r_{Swave}($GeV$^{-1})$ & 0 -- fixed         & 0 -- fixed         & 0 -- fixed    \\
Other $R \to K\pi$	                            
     & $r_R($GeV$^{-1})$       & 1.5 -- fixed       & 1.5 -- fixed       & 1.5 -- fixed  \\
\hline			                            
     & $\sum$ FF$_i$ (\%)      & 94.4               & 152.0              & 122.8         \\
Goodness		                            
     & $\chi^2/\nu$            & 426/385            & 416/385            & 359/347       \\  
     & Probability (\%)        & 7.4\%              & 13.2\%             & 31.5\%        \\  
\hline
\hline
\end{tabular}
\end{center}
\end{table}

\section{Quasi-model-independent partial wave analysis}
\label{sec:QMIPWA_I2}
\subsection{\boldmath QMIPWA for $K\pi$ $S$ wave}
\label{sec:QMIPWA_for_Swave}

The biggest issue of any Dalitz plot analysis is its model dependence.
An attempt to mitigate the model dependence for the decay under study
is described in \cite{E791_Kpipi}.
Here we reproduce this analysis using a slightly modified technique,
which we call the quasi-model-independent partial wave analysis (QMIPWA).
We apply this technique as an extension of our model~I2.

In QMIPWA we modify the parameters of model~I2 as follows.
The $NR$ and $\kappa\pi$ components in model~I2
are replaced by the binned $S$ wave amplitude [Eq.~(\ref{eqn:binned_wave})].
The 26 binned $S$ wave magnitudes and phases
of $m^2_{K\pi}$ are floating fit parameters.
We keep the $K_0^*(1430)\pi$ contribution in its Breit-Wigner form,
because it has a sharp structure, that cannot be well reproduced by
a binned amplitude.
The $K_0^*(1430)$ parameters are fixed to their values from model~I2 in order to remove
correlations between the Breit-Wigner function and the binned $S$ wave amplitude.
The $K\pi$ $P$ wave in Eq.~(\ref{eqn:A_1}) 
is represented by the $K^*(892)$ and $K^*(1680)$ Breit-Wigner functions,
where the $c_{K^*(1680)}$ parameters are allowed to float in the fit.
As usual $c_{K^*(892)}=1$, and all other magnitudes and phases 
are defined with respect to $K^*(892)$.
The $K^*(1410)$ is excluded as mentioned previously.
The $K\pi$ $D$ wave in Eq.~(\ref{eqn:A_2}) 
is represented by the $K_2^*(1430)$ Breit-Wigner function with its
parameters $c_{K_2^*(1430)}$ allowed to float in the fit.
The $I=2$ $\pi^+\pi^+$ $S$ wave is represented by the unitary amplitude 
from Eq.~(\ref{eqn:I2pipiSwave}) fixing the
$m_{min}$ and $m_{max}$ parameters to their optimal values from model~I2.
The $a_{\pi\pi}$ and $\phi_{\pi\pi}$ parameters are allowed to float in the fit.

\begin{figure}[!htb]
  \includegraphics[width=160mm]{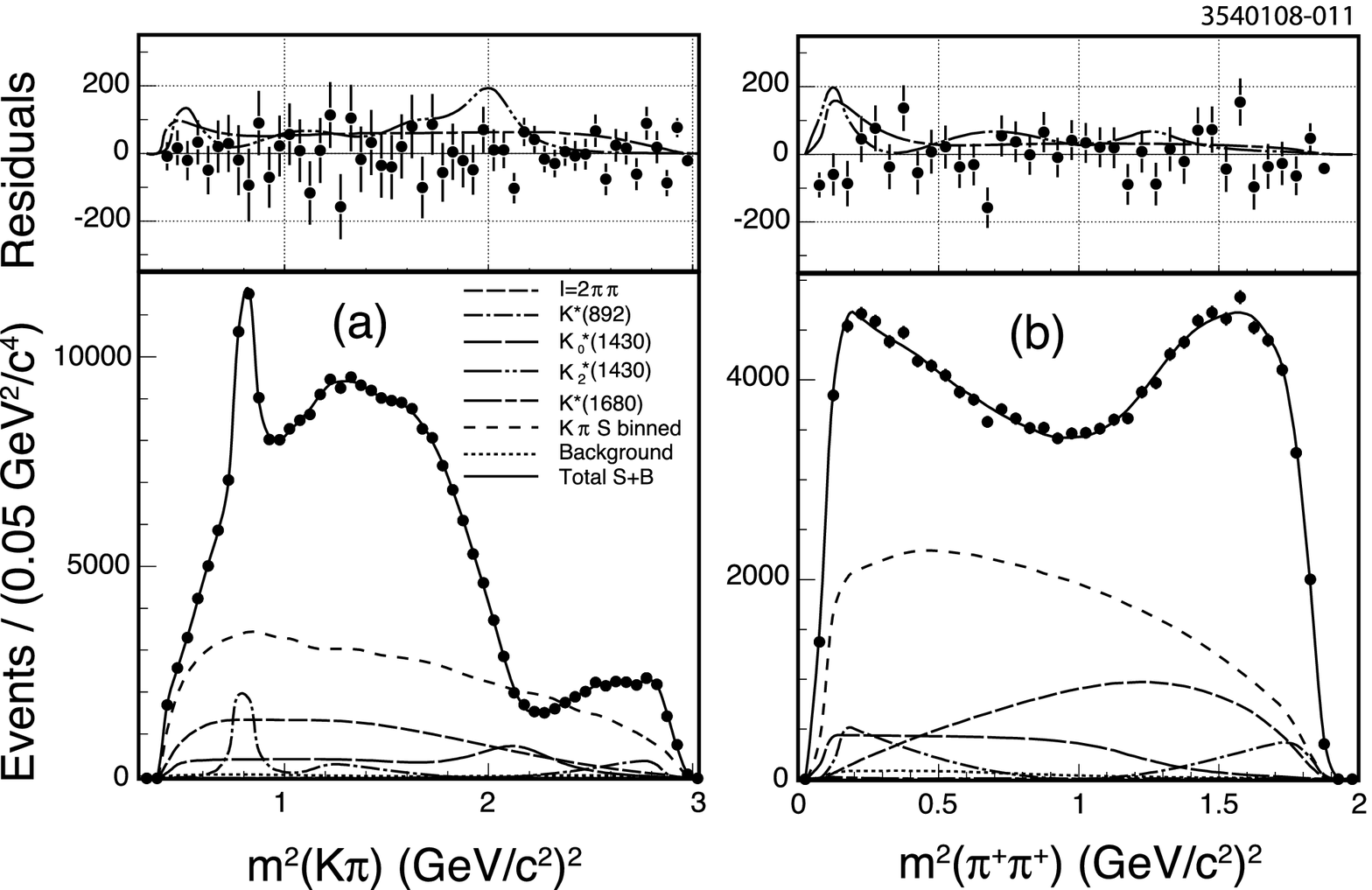}
  \caption{\label{fig:Projections_Model_QMIPWA_I2} 
                                             Projections of the fit to the Dalitz plot with QMIPWA on 
                                         (a) $m^2(K\pi)$ (two entries per event) and 
                                         (b) $m^2(\pi\pi)$ variables. Residuals between data and
                                         the total p.d.f. are shown by dots 
                                         with statistical error bars on the
                                         top inset along with minor contributions from the 
                                         $K^*(1680)$ and $K_2^*(1430)$ resonances, plotted with factor 
                                         $\times$10.} 
\end{figure}

The results of the QMIPWA fit are shown in Table~\ref{tab:Isobar_I2_vs_QMIPWA_I2} 
for resonance parameters and
Table~\ref{tab:results_Swave_with_I2} for the $K\pi$ $S$ wave.
Figure~\ref{fig:Projections_Model_QMIPWA_I2} shows the Dalitz plot projections of this fit.
The measured $K\pi$ $S$ wave and their comparison with model~I2 components 
are displayed in Fig.~\ref{fig:Swave_QMIPWA_I2}.
Both the magnitude and phase are different from those of model~I2.

Table~\ref{tab:results_Swave_with_I2} (right-hand side) also shows 
the result of a similar fit for a total binned $S$ wave. 
For this fit the $K_0^*(1430)$ resonance is accounted for in the binned $S$ wave,
and all other parameters are fixed to their values from the nominal QMIPWA fit.

\subsection{\boldmath Cross-check for binned $I=2$ $\pi^+\pi^+$ $S$ wave}
\label{sec:binned_I2}

In this approach we also check how much the $I=2$ $\pi^+\pi^+$ $S$ wave 
might be different from its analytic approximation in model~I2.
The analytic expression, given in Eq.~(\ref{eqn:I2pipiSwave}), 
is replaced by the binned wave from Eq.~(\ref{eqn:binned_wave}).
We fix all 20 fit parameters to their values in model~I2.
Then, we fix to zero the magnitude of the analytic $I=2$ $\pi^+\pi^+$ $S$ wave,
$a_{\pi\pi}=0$, 
and add a binned $I=2$ $\pi^+\pi^+$ $S$ wave; the  
magnitude and phase in each of the 18 bins of $m^2_{\pi^+\pi^+}$ are allowed to float.
Results of this fit are shown in Table~\ref{tab:results_I2_pipi_Swave}.
The measured $I=2$ $\pi^+\pi^+$ $S$ wave is compared to the
model~I2 analytic function in Fig.~\ref{fig:pipiSwave}.
The change of the $\chi^2$ in this fit compared to model~I2,
$\Delta\chi^2 = 390-416 = -26$, 
does not show a significant improvement for the change 
$\Delta\nu = -36+20 = -16$ in degrees of freedom due to
fixing 20 of the original parameters and introducing 36 new parameters.
The data do not prefer the binned amplitude to the analytic model
for the $I=2$ $\pi^+\pi^+$ $S$ wave.

\subsection{\boldmath Cross-checks for binned $K\pi$ $P$ and $D$ waves}
\label{sec:binned_P_and_D}

To check how much the $P$ and $D$ waves might be different 
from their model parametrization we use the same binned technique.
All of the fit parameters are fixed to their optimal values from model~I2.
The $P$ wave binned amplitude substitutes for the smooth $K^*(1680)$ resonance only.
The sharp shape of the $K^*(892)$ resonance is accounted for by a Breit-Wigner function.
The $D$ wave binned amplitude substitutes for the $K_2^*(1430)$ contribution, even
though it is not particularly smooth.
The magnitude and phase in 26 bins of $m^2_{K\pi}$ 
are allowed to float for the $P$ or $D$ wave in two separate fits, respectively.
The resolution for the $D$ wave is worse, and we use only 18 bins in the
$m^2_{K\pi}$ range [0.9,~2.7]~(GeV/$c^2$)$^2$.
Results of these fits are shown in Table~\ref{tab:results_PandD_waves}.
The measured $P$ and $D$ binned waves and their comparison with the model~I2 components
are displayed in Figs.~\ref{fig:Pwave} and \ref{fig:Dwave}, respectively.

The relative fraction of the $P$ and especially the $D$ wave is much smaller than the $S$ wave.
This explains the poor resolution for these binned amplitudes.
We find that our binned $P$ and $D$ waves are consistent
with their substituted components in model~I2. 
This cross-check adds some confidence to this quasi-model-independent technique.

\begin{table}[!htb]
\caption{\label{tab:results_Swave_with_I2} QMIPWA: Results for $K\pi$ $S$ wave. 
         Figure~\ref{fig:Swave_QMIPWA_I2} shows the binned $K\pi$ $S$ wave without $K_0^*(1430)$
         by the dots with error bars and the total $K\pi$ $S$ wave by the solid curve. 
        }
\begin{center}
\begin{tabular}{c c c c c c }
\hline
\hline
Bin & \multicolumn{1}{|c|}{$m^2_{K\pi}$ (GeV/$c^2$)$^2$} 
    & \multicolumn{2}{ c|}{Binned $K\pi$ $S$ wave w/o $K_0^*(1430)$}
    & \multicolumn{2}{ c }{Total $K\pi$ $S$ wave} \\  
      \cline{3-6}
\#
& \multicolumn{1}{|c|}{ Bin range }   
& \multicolumn{1}{ c|}{ Magnitude (a.u.) }  
& \multicolumn{1}{ c|}{ Phase ($^\circ$) }               
& \multicolumn{1}{ c|}{ Magnitude (a.u.) }
& \multicolumn{1}{ c }{ Phase ($^\circ$) } \\
\hline
  1 &  0.4 --- 0.5 & 20.23$\pm$0.80$\pm$0.51 &--71.8$\pm$4.9$\pm$4.1 &  19.29$\pm$0.56 &--66.0$\pm$3.2 \\
  2 &  0.5 --- 0.6 & 20.90$\pm$0.68$\pm$0.54 &--61.1$\pm$4.6$\pm$2.3 &  20.38$\pm$0.51 &--54.3$\pm$2.8 \\
  3 &  0.6 --- 0.7 & 20.58$\pm$0.63$\pm$0.69 &--48.8$\pm$4.0$\pm$3.2 &  20.35$\pm$0.45 &--40.9$\pm$2.2 \\
  4 &  0.7 --- 0.8 & 20.62$\pm$0.64$\pm$0.51 &--47.0$\pm$2.4$\pm$0.7 &  20.52$\pm$0.59 &--39.7$\pm$0.9 \\
  5 &  0.8 --- 0.9 & 20.95$\pm$0.74$\pm$1.32 &--44.1$\pm$3.6$\pm$3.5 &  20.84$\pm$0.49 &--36.0$\pm$1.4 \\
  6 &  0.9 --- 1.0 & 19.97$\pm$0.63$\pm$1.14 &--38.4$\pm$4.0$\pm$6.4 &  19.97$\pm$0.36 &--28.7$\pm$2.7 \\
  7 &  1.0 --- 1.1 & 19.36$\pm$0.58$\pm$0.96 &--26.2$\pm$3.5$\pm$4.8 &  19.97$\pm$0.34 &--15.7$\pm$2.6 \\
  8 &  1.1 --- 1.2 & 17.81$\pm$0.53$\pm$0.96 &--18.6$\pm$3.0$\pm$6.1 &  18.85$\pm$0.30 & --6.9$\pm$2.0 \\
  9 &  1.2 --- 1.3 & 17.70$\pm$0.49$\pm$0.90 &--13.7$\pm$2.6$\pm$5.1 &  19.05$\pm$0.27 & --1.2$\pm$1.6 \\
 10 &  1.3 --- 1.4 & 17.72$\pm$0.47$\pm$1.05 & --8.5$\pm$2.5$\pm$5.8 &  19.47$\pm$0.25 &   4.8$\pm$1.4 \\
 11 &  1.4 --- 1.5 & 17.13$\pm$0.45$\pm$1.05 & --3.2$\pm$2.4$\pm$4.7 &  19.36$\pm$0.23 &  11.6$\pm$1.3 \\
 12 &  1.5 --- 1.6 & 17.16$\pm$0.45$\pm$1.04 &   0.5$\pm$2.3$\pm$5.6 &  19.80$\pm$0.22 &  17.1$\pm$1.3 \\
 13 &  1.6 --- 1.7 & 17.09$\pm$0.46$\pm$1.14 &   5.4$\pm$2.3$\pm$4.0 &  20.35$\pm$0.22 &  24.3$\pm$1.3 \\
 14 &  1.7 --- 1.8 & 16.94$\pm$0.47$\pm$1.06 &   7.5$\pm$2.3$\pm$4.6 &  20.61$\pm$0.24 &  30.1$\pm$1.3 \\
 15 &  1.8 --- 1.9 & 16.41$\pm$0.48$\pm$1.05 &   9.5$\pm$2.4$\pm$3.7 &  20.38$\pm$0.27 &  38.0$\pm$1.3 \\
 16 &  1.9 --- 2.0 & 15.91$\pm$0.51$\pm$0.75 &  12.3$\pm$2.7$\pm$3.2 &  19.86$\pm$0.34 &  49.1$\pm$1.4 \\
 17 &  2.0 --- 2.1 & 15.97$\pm$0.56$\pm$1.11 &  16.9$\pm$2.9$\pm$3.1 &  19.07$\pm$0.40 &  63.4$\pm$1.4 \\
 18 &  2.1 --- 2.2 & 15.72$\pm$0.61$\pm$1.11 &  15.7$\pm$2.5$\pm$2.2 &  14.20$\pm$0.40 &  76.6$\pm$1.9 \\
 19 &  2.2 --- 2.3 & 16.54$\pm$0.63$\pm$1.17 &  17.3$\pm$2.1$\pm$2.1 &   9.29$\pm$0.39 &  75.2$\pm$2.6 \\
 20 &  2.3 --- 2.4 & 16.64$\pm$0.71$\pm$1.42 &  20.1$\pm$2.1$\pm$3.4 &   7.10$\pm$0.43 &  57.5$\pm$3.1 \\
 21 &  2.4 --- 2.5 & 15.91$\pm$0.86$\pm$1.61 &  19.8$\pm$2.3$\pm$2.0 &   6.75$\pm$0.43 &  33.5$\pm$3.2 \\
 22 &  2.5 --- 2.6 & 17.25$\pm$1.03$\pm$1.83 &  21.5$\pm$2.4$\pm$2.0 &   9.46$\pm$0.51 &  26.0$\pm$2.5 \\
 23 &  2.6 --- 2.7 & 17.24$\pm$1.21$\pm$2.38 &  24.1$\pm$2.6$\pm$2.9 &  10.74$\pm$0.55 &  26.7$\pm$2.6 \\
 24 &  2.7 --- 2.8 & 17.59$\pm$1.34$\pm$2.28 &  29.0$\pm$2.8$\pm$1.6 &  12.04$\pm$0.64 &  31.8$\pm$2.5 \\
 25 &  2.8 --- 2.9 & 16.51$\pm$1.65$\pm$2.56 &  32.7$\pm$3.5$\pm$3.5 &  11.79$\pm$0.79 &  36.0$\pm$3.3 \\
 26 &  2.9 --- 3.0 & 14.07$\pm$3.28$\pm$3.32 &  35.6$\pm$6.0$\pm$3.8 &   9.75$\pm$2.50 &  38.6$\pm$6.1 \\
\hline
\hline
\end{tabular}
\end{center}
\end{table}

\begin{table}[!htb]
\caption{\label{tab:results_I2_pipi_Swave} QMIPWA: Results for $\pi^+\pi^+$ $S$ wave, 
         which are also shown by dots with error bars in Fig.~\ref{fig:pipiSwave}.
	 Variation of the $\chi^2$ and number of degrees of freedom, $\Delta\nu$,  
         is shown with respect to model~I2. 
        }
\begin{center}
\begin{tabular}{ c c c c }
\hline
\hline
Bin 
& \multicolumn{1}{|c}{ $m^2_{\pi\pi}$ (GeV/$c^2$)$^2$ }
& \multicolumn{2}{|c}{ $I=2$ $\pi^+\pi^+$ $S$ wave } \\  
      \cline{3-4}
\#  
& \multicolumn{1}{|c}{ Bin range }  
& \multicolumn{1}{|c}{ Magnitude (a.u.) }
& \multicolumn{1}{|c}{ Phase ($^\circ$) } \\
\hline
  1 & 0.1 --- 0.2 &    3.62$\pm$0.44 & --113.9$\pm$7.0  \\
  2 & 0.2 --- 0.3 &    4.31$\pm$0.37 & --115.8$\pm$4.7	\\
  3 & 0.3 --- 0.4 &    4.51$\pm$0.33 & --120.7$\pm$3.6	\\
  4 & 0.4 --- 0.5 &    6.17$\pm$0.33 & --116.8$\pm$2.4	\\
  5 & 0.5 --- 0.6 &    6.84$\pm$0.34 & --118.9$\pm$2.1	\\
  6 & 0.6 --- 0.7 &    8.13$\pm$0.35 & --121.8$\pm$1.8	\\
  7 & 0.7 --- 0.8 &    7.77$\pm$0.35 & --121.7$\pm$1.9	\\
  8 & 0.8 --- 0.9 &    8.65$\pm$0.35 & --125.5$\pm$1.7	\\
  9 & 0.9 --- 1.0 &    8.95$\pm$0.37 & --126.3$\pm$1.6	\\
 10 & 1.0 --- 1.1 &    9.61$\pm$0.42 & --127.9$\pm$1.5	\\
 11 & 1.1 --- 1.2 &   11.69$\pm$0.55 & --127.9$\pm$1.3	\\
 12 & 1.2 --- 1.3 &   10.04$\pm$0.93 & --132.0$\pm$1.5	\\
 13 & 1.3 --- 1.4 &   12.43$\pm$1.01 & --131.0$\pm$1.2	\\
 14 & 1.4 --- 1.5 &   12.92$\pm$0.98 & --131.1$\pm$1.3	\\
 15 & 1.5 --- 1.6 &    9.51$\pm$0.98 & --131.4$\pm$1.8	\\
 16 & 1.6 --- 1.7 &   11.56$\pm$1.00 & --134.7$\pm$1.8	\\
 17 & 1.7 --- 1.8 &   11.75$\pm$1.10 & --142.5$\pm$2.1	\\
 18 & 1.8 --- 1.9 &   10.52$\pm$2.02 & --156.1$\pm$5.2	\\
\hline
\multicolumn{2}{c}{FF(\%)}         & 
\multicolumn{2}{c}{9.8$\pm$0.4}                         \\
\multicolumn{2}{c}{$\Delta\chi^2 / \Delta\nu$ } &
\multicolumn{2}{c}{$(390-416)/(-36+20)$}                 \\
\hline
\hline
\end{tabular}
\end{center}
\end{table}

\begin{table}[!htb]
\caption{\label{tab:results_PandD_waves} Results for $K\pi$ $P$ and $D$ binned waves
         using model~I2, which is also shown by dots with error bars in  
         Figs.~\ref{fig:Pwave} and
               \ref{fig:Dwave}, respectively.
	 Variation of the $\chi^2$ and number of degrees of freedom, $\Delta\nu$, 
         is shown with respect to model~I2.   The $D$ wave is used only in the range
         [0.9,2.7] (GeV/$c^2$)$^2$ as explained in the text.
        }
\begin{center}
\begin{tabular}{ c c c c c c}
\hline
\hline
Bin 
& \multicolumn{1}{|c|}{ $m^2_{K\pi}$ (GeV/$c^2$)$^2$ }  
& \multicolumn{2}{ c|}{ Binned $P$ wave for $K^*(1680)$ } 
& \multicolumn{2}{ c }{ Binned $D$ wave for $K_2^*(1430)$} \\  
      \cline{3-6}
\#  
& \multicolumn{1}{|c|}{ Bin range        }
& \multicolumn{1}{ c|}{ Magnitude (a.u.) }     
& \multicolumn{1}{ c|}{ Phase ($^\circ$) }  
& \multicolumn{1}{ c|}{ Magnitude (a.u.) }     
& \multicolumn{1}{ c }{ Phase ($^\circ$) }       \\
\hline 
  1 & 0.4 --- 0.5 &   1.96$\pm$1.26 &  41.0$\pm$28.0 &     $\cdots$         & $\cdots$     \\
  2 & 0.5 --- 0.6 &   2.90$\pm$1.12 & 199.2$\pm$11.2 &     $\cdots$         & $\cdots$     \\
  3 & 0.6 --- 0.7 &   0.67$\pm$0.93 &  50.1$\pm$15.8 &     $\cdots$         & $\cdots$     \\
  4 & 0.7 --- 0.8 &   0.81$\pm$0.36 &  46.4$\pm$18.4 &     $\cdots$         & $\cdots$     \\
  5 & 0.8 --- 0.9 &   0.60$\pm$0.35 & 102.3$\pm$33.7 &     $\cdots$         & $\cdots$     \\
  6 & 0.9 --- 1.0 &   0.58$\pm$1.36 &  73.2$\pm$25.7 &      0.71$\pm$0.17 &--61.8$\pm$23.6 \\
  7 & 1.0 --- 1.1 &   1.87$\pm$0.77 &  79.5$\pm$7.01 &      0.31$\pm$0.14 & --5.0$\pm$69.6 \\
  8 & 1.1 --- 1.2 &   0.09$\pm$0.42 &  51.7$\pm$22.4 &      0.66$\pm$0.16 &--15.6$\pm$27.2 \\
  9 & 1.2 --- 1.3 &   1.12$\pm$0.38 &  81.3$\pm$10.6 &      1.14$\pm$0.19 &--50.0$\pm$12.9 \\
 10 & 1.3 --- 1.4 &   1.49$\pm$0.33 &  74.4$\pm$8.79 &      1.09$\pm$0.19 &--28.7$\pm$13.1 \\
 11 & 1.4 --- 1.5 &   0.81$\pm$0.33 &  83.9$\pm$13.3 &      0.97$\pm$0.19 & --9.9$\pm$14.2 \\
 12 & 1.5 --- 1.6 &   1.25$\pm$0.33 &  80.8$\pm$10.6 &      1.30$\pm$0.21 &--15.5$\pm$14.0 \\
 13 & 1.6 --- 1.7 &   1.30$\pm$0.34 &  73.3$\pm$9.79 &      1.45$\pm$0.24 &  11.6$\pm$13.5 \\
 14 & 1.7 --- 1.8 &   1.77$\pm$0.36 &  77.7$\pm$8.34 &      2.18$\pm$0.26 &--16.3$\pm$10.2 \\
 15 & 1.8 --- 1.9 &   1.16$\pm$0.40 &  88.1$\pm$11.1 &      3.28$\pm$0.28 &  24.4$\pm$7.83 \\
 16 & 1.9 --- 2.0 &   1.79$\pm$0.45 &  90.9$\pm$9.91 &      5.77$\pm$0.38 &  31.9$\pm$5.45 \\
 17 & 2.0 --- 2.1 &   1.86$\pm$0.42 & 102.1$\pm$11.2 &      4.77$\pm$0.52 &  70.2$\pm$5.90 \\
 18 & 2.1 --- 2.2 &   2.42$\pm$0.45 & 104.3$\pm$12.2 &      4.45$\pm$0.70 & 100.2$\pm$7.41 \\
 19 & 2.2 --- 2.3 &   2.58$\pm$0.52 & 115.7$\pm$12.6 &      4.69$\pm$0.97 & 127.8$\pm$8.52 \\
 20 & 2.3 --- 2.4 &   1.66$\pm$0.55 & 119.6$\pm$16.6 &      3.79$\pm$1.13 & 116.8$\pm$16.3 \\
 21 & 2.4 --- 2.5 &   3.76$\pm$0.60 & 108.8$\pm$11.4 &      0.18$\pm$1.42 & --3.3$\pm$84.7 \\
 22 & 2.5 --- 2.6 &   4.19$\pm$0.75 & 113.6$\pm$13.7 &      4.86$\pm$1.72 & 135.8$\pm$33.2 \\
 23 & 2.6 --- 2.7 &   6.71$\pm$0.98 & 105.0$\pm$16.9 &      1.54$\pm$2.58 & 184.8$\pm$78.5 \\
 24 & 2.7 --- 2.8 &   3.35$\pm$1.59 & 142.4$\pm$29.4 &     $\cdots$ 	  & $\cdots$       \\
 25 & 2.8 --- 2.9 &   7.03$\pm$2.49 & 183.2$\pm$22.5 &     $\cdots$ 	  & $\cdots$       \\
 26 & 2.9 --- 3.0 &  32.66$\pm$17.9 & 232.2$\pm$11.6 &     $\cdots$ 	  & $\cdots$       \\
\hline									     
\multicolumn{2}{c|}{FF(\%), 2$\times$} & 
\multicolumn{2}{c|}{0.20$\pm$0.03}     & 
\multicolumn{2}{c }{0.15$\pm$0.01}     \\
\multicolumn{2}{c|}{$\Delta\chi^2 / \Delta\nu$}   &			     
\multicolumn{2}{c|}{$(373-416) / (-52+20)$ }       & 
\multicolumn{2}{c }{$(400-416) / (-36+20)$ }       \\
\hline
\hline
\end{tabular}
\end{center}
\end{table}

\section{Cross-checks and systematic uncertainties}
\label{sec:systematic_cross_checks}

\subsection{Systematic uncertainties}
\label{sec:systematic_uncertainties}

In order to estimate systematic uncertainties of the fit parameters,
we apply numerous variations to the fitting procedure and examine 
the change of the fit parameters.
Because of the numerous variations, a quadratic sum of the resulting
small  changes due to the variations would lead to a systematic error that
is too conservative.  Instead, we treat all of the resulting changes
from the variations as a frequency distribution and calculate the
{\it mean} and {\it rms} of the distribution. 
The {\it total} systematic uncertainty is obtained as a
quadratic sum of the difference between the obtained {\it mean} and the default
value of that parameter, and the {\it rms}.

We consider the six divisions of our sample:
splitting events evenly between earlier and later datasets;
$D^+$ and $D^-$ decays; 
tight [$1\sigma(\Delta E) \times 1\sigma(m_{\rm BC})$];
and loose [$3\sigma(\Delta E) \times 3\sigma(m_{\rm BC})$] signal box cuts.
Other systematic variations common to all of the models 
(model~C, model~I2, and QMIPWA), are as follows: 
\begin{itemize}
\item            Float $f_{\rm sig}$.  We find that it is always consistent with
            its nominal value of 0.9892. 
\item
            Drop the background term and set $f_{\rm sig}=1$.
\item            Float the efficiency coefficients in a simultaneous fit to data and MC samples. 
            In this case variations of the efficiency parameters are constrained by the 
            MC sample.
\item            Improve the precision of the calculation of the normalization integrals from Eq.~(\ref{eqn:PDF}) 
            by an order of magnitude.
\end{itemize}
We repeated the fits with the radii for the Blatt-Weisskopf~\cite{Blatt-Weisskopf} form factors
a factor of 2 larger and smaller than their nominal values and found
negligible change in the central result.

Depending on the model we apply additional systematic variations, 
which are also included in systematic uncertainties.
In model~C and model~I2, discussed in Sec.~\ref{sec:isobar_model}, 
we consider variation of the resonance parameters as follows:
\begin{itemize}
\item       Float parameters of the $K^*(892)$ and $K_0^*(1430)$ resonances. 
            The mass and width are shown in Table~\ref{tab:model_parameters}. 
\item
            Float parameters of the $K_2^*(1430)$ or $K^*(1680)$ resonances. 
            The mass and width are shown in Table~\ref{tab:model_parameters}.
\item       Add a $K^*(1410)$ contribution.  We find that it is not significant:
            $FF_{K^*(1410)}<0.1\%$ at 90\% C.L.
\item
  In model C we test the event sample selected without a kinematic fit 
  with appropriate coefficients for efficiency and background shapes.
We find that the resulting variation is negligible, and this study
    is not performed for the other fit models.
\item
  In QMIPWA we use the binned $K^-\pi^+$ $S$ wave as a histogram without linear interpolation.
\end{itemize}
None of these variations reveal an obviously dominant source of systematic
uncertainty.

\subsection{Additional cross-checks}
\label{sec:cross_checks}

Common cross-checks listed below
bring us to models that are different enough that we do not include these results 
in the total systematic uncertainty.
\begin{itemize}
\item[$\circ$]
            Remove the $K^*_2(1430)$ resonance contribution.  We find 
            a significant degradation of the fit quality.
            For example, $\Delta\chi^2 = 370$ in model~C.
\item[$\circ$] 
            Remove the $K^*(1680)$ resonance contribution.  We find
            a significant degradation of the fit quality.
            For example, $\Delta\chi^2 = 250$ in model~C.
\end{itemize}
  Based on these fits we conclude that
  the $K^*_2(1430)$ and $K^*(1680)$ resonances cannot be removed. 
  Their contributions are small (FF$<1\%$) but significant and the
  fit quality is very poor without these resonances.
  We apply a similar backward check for model~I2 and QMIPWA:
\begin{itemize}
\item[$\circ$]
  We remove the $I=2$ $\pi^+\pi^+$ $S$ wave contribution and get a poor quality fit.  
\end{itemize}

\begin{figure}[!htb]
  \includegraphics[width=100mm]{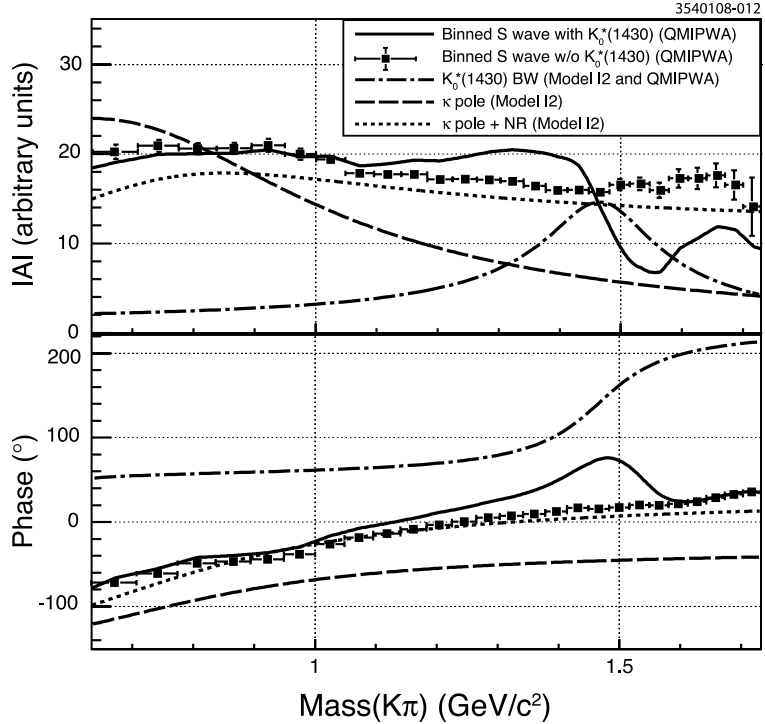}
  \caption{\label{fig:Swave_QMIPWA_I2} The magnitude and phase of the
                                       $K\pi$ $S$ wave in model~I2 and QMIPWA.
                                       The dots with error bars for statistical uncertainties
                                       and the solid curve  
                                       show the binned $K\pi$ $S$ wave without $K_0^*(1430)$
                                       and the total $K\pi$ $S$ wave from
                                       Table~\ref{tab:results_Swave_with_I2}, respectively.
                                       Other curves show the $S$ wave components of model~I2 
                                       with parameters from 
                                       Table~\ref{tab:Isobar_I2_vs_QMIPWA_I2}.} 
\end{figure}

\begin{figure}[!htb]
  \includegraphics[width=100mm]{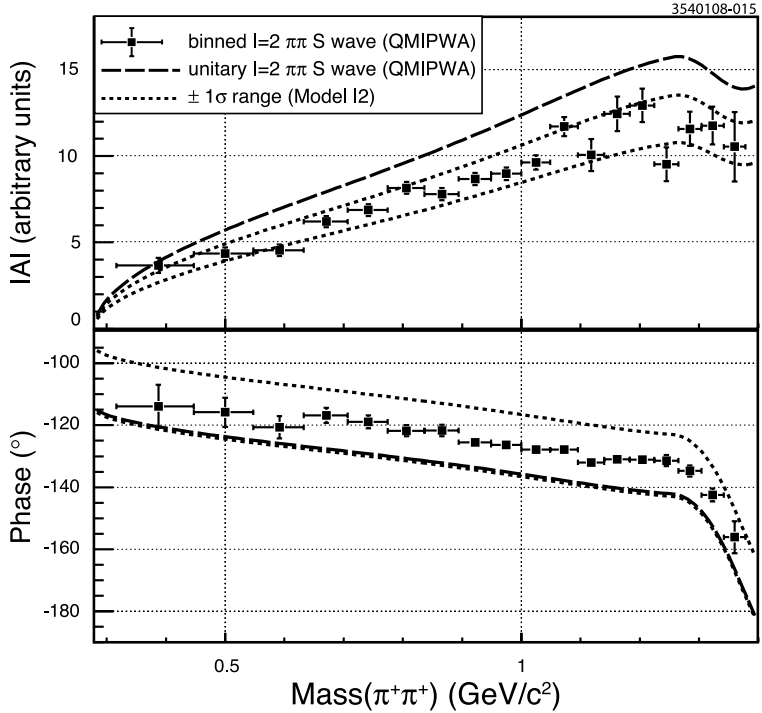}
  \caption{\label{fig:pipiSwave} The magnitude and phase of the $I=2$ $\pi^+\pi^+$ $S$ wave 
                                 in model~I2 and QMIPWA.
                                 The dots with error bars for statistical uncertainties
                                 represent results from 
                                 Table~\ref{tab:results_I2_pipi_Swave}.
                                 Other curves show the $I=2$ $\pi^+\pi^+$ $S$ wave of the model~I2 
                                 and QMIPWA with parameters from 
                                 Table~\ref{tab:Isobar_I2_vs_QMIPWA_I2}.} 
\end{figure}

\begin{figure}[!htb]
  \includegraphics[width=100mm]{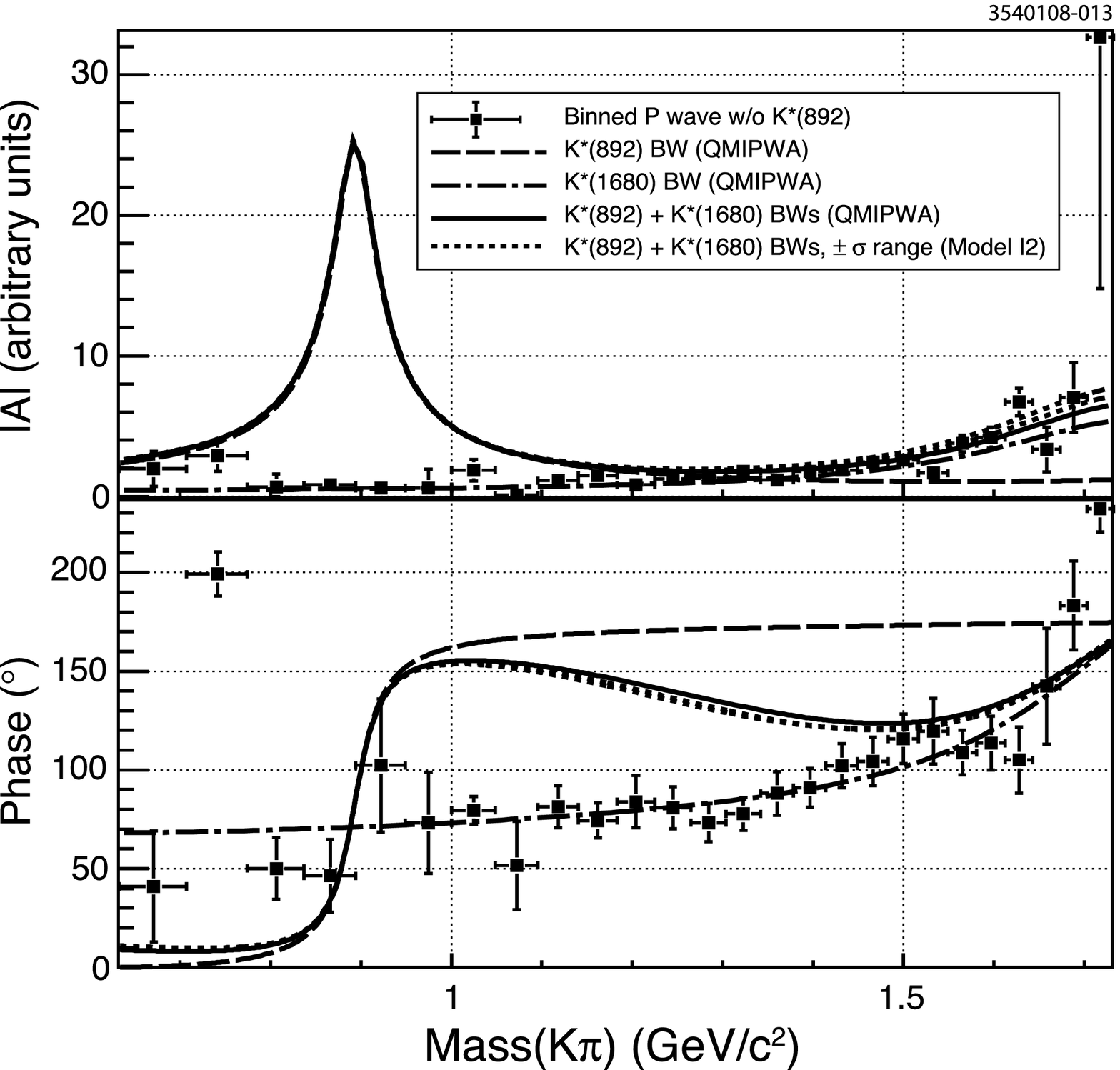}
  \caption{\label{fig:Pwave} The magnitude and phase of the $K\pi$ $P$ wave 
                             in model~I2 and QMIPWA.
                             The dots with error bars for statistical uncertainties
                             represent results from 
                             Table~\ref{tab:results_PandD_waves}.
                             Other curves show the $P$ wave components of model~I2  
                             and QMIPWA with parameters from 
                             Table~\ref{tab:Isobar_I2_vs_QMIPWA_I2}.} 
\end{figure}

\begin{figure}[!htb]
  \includegraphics[width=100mm]{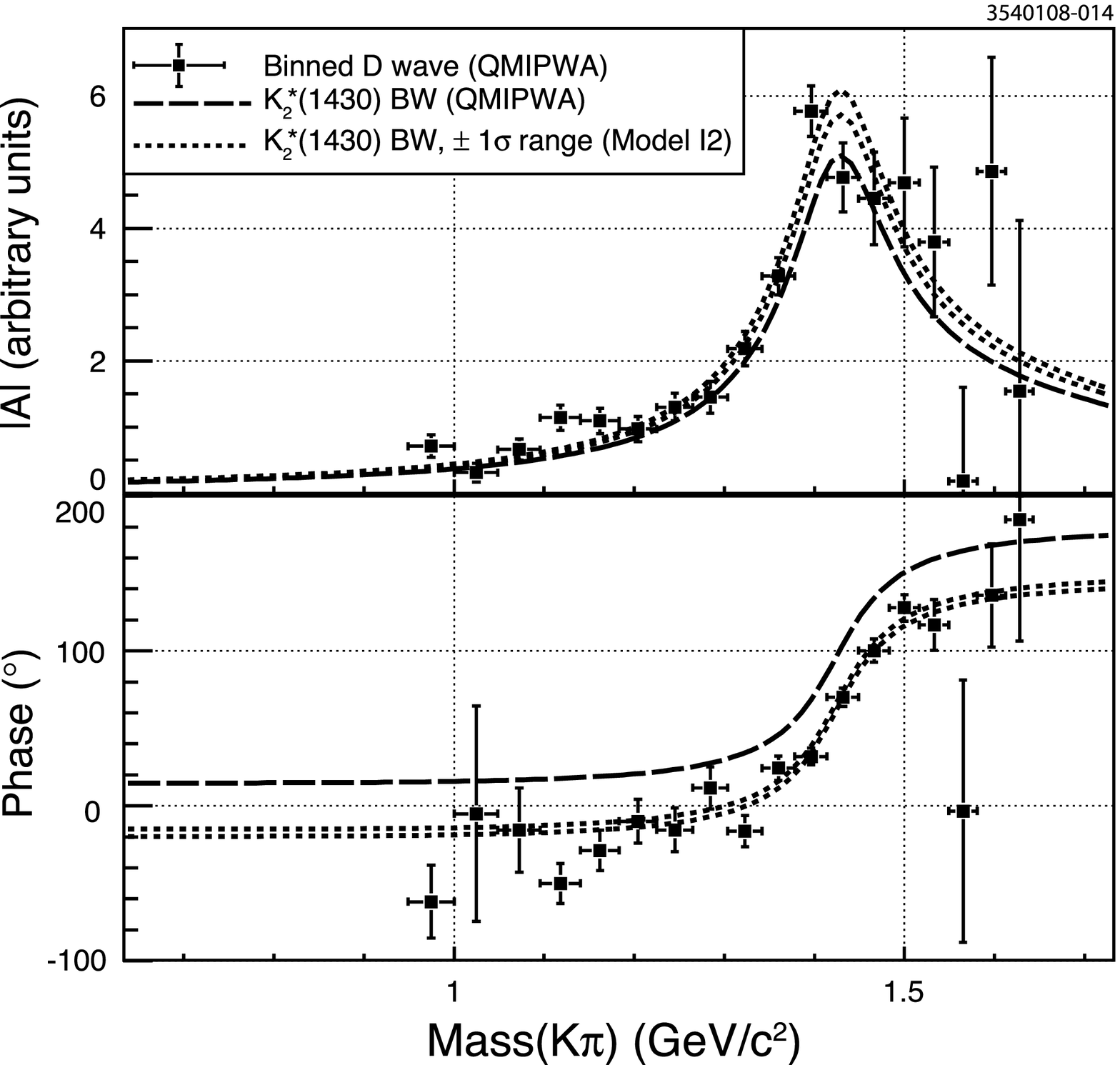}
  \caption{\label{fig:Dwave} The magnitude and phase of the $K\pi$ $D$ wave 
                             in model~I2 and QMIPWA.
                             The dots with error bars for statistical uncertainties 
                             represent results from 
                             Table~\ref{tab:results_PandD_waves}.
                             Other curves show the $D$ wave components of model~I2  
                             and QMIPWA with parameters from 
                             Table~\ref{tab:Isobar_I2_vs_QMIPWA_I2}.} 
\end{figure}

\section{Discussion}
\label{sec:discussion}

\subsection{\boldmath $K_0^*(1430)$ and $K^*(892)$ parameters}
\label{sec:resonance_parameters}

  The $K_0^*(1430)$ parameters, which we obtain in fits with the isobar
  models, listed in Table~\ref{tab:model_parameters} as well as in 
  Tables~\ref{tab:modelC}-\ref{tab:Isobar_I2_vs_QMIPWA_I2}, 
  are consistent with each other and are significantly
  different from the PDG~\cite{PDG-2006} values which are dominated by the LASS measurement.
  Our data prefer a $K_0^*(1430)$ resonance that is 
  50~MeV/$c^2$ heavier and about 2 times narrower. 
  Our result is consistent with E791~\cite{E791-2002} 
  and FOCUS~\cite{FOCUS-2007} measurements.
  Possible explanations include that
  the $K_0^*(1430)$ resonance parameters depend on its production mechanism 
  which are different in $K\pi$ scattering and $D$-meson decays or
  that the $K_0^*(1430)$ parameters
  are strongly model-dependent. In particular, they may depend on interference with 
  other $S$ wave contributions.

  In all of our fits, as seen in 
  Tables~\ref{tab:model_parameters} and 
         \ref{tab:Isobar_I2_vs_QMIPWA_I2}, the $K^*(892)$
  parameters are consistent with each other. The measured masses are consistent
  with the nominal value~\cite{PDG-2006}, but the widths are about 5 MeV/$c^2$ narrower.
  The $K^*(892)$ width in the PDG 
  is an average value over about 20 experimental measurements.
  Our measurement does not contradict any single
  result included in PDG. The largest difference is about 3 standard deviations
  from the LASS value. 
  We also tested the $K^*(892)$ width model dependence. 
  The largest variation is expected from applying Blatt-Weisskopf form factors. 
  A variation of the radial parameters from zero 
  to 4 times their nominal values causes a variation in the width
  of 0.7~MeV/$c^2$, which cannot explain the large difference with the PDG average.
  Recent results from FOCUS~\cite{FOCUS-Kst-width} and 
  Belle~\cite{Belle-Kst-width} also indicate a $K^*(892)$ width
  smaller than the PDG value. 

\subsection{Partial waves}
\label{sec:discussion_partial_waves}

  The factorized Gaussian form factors used in model~C for the $K\pi$ $S$ wave components
  may cause an enhancement of the complex function magnitude at low $K\pi$ mass, 
  as mentioned in Sec.~\ref{sec:variations_of_modelC}.
  However, the form factor is a real function and does not change the
  phase of the complex $S$ wave. 
  In model~I2 and QMIPWA we measure the total $K\pi$ $S$ wave amplitude 
  without using a form factor. It means that the measured $S$ wave absorbs the form factor.
  The measured $S$ wave magnitude is essentially constant up to 1.4~GeV/$c^2$,
  as demonstrated in Fig.~\ref{fig:Swave_QMIPWA_I2}.
  The phase of the complex $S$ wave amplitude
  shows smooth variation from --80$^\circ$
  at $K\pi$ threshold to 40$^\circ$ at 1.4~GeV/$c^2$.  
  At higher mass the amplitude is distorted by a
  contribution from the $K_0^*(1430)$ resonance. 
  Our data require a dominant contribution from the $K\pi$ $S$ wave, 
  which at low mass is not well described as a regular
  resonance structure.

  In QMIPWA the $S$ wave is measured in a model-independent way,
  while the $P$ and $D$ waves are parametrized by Breit-Wigner resonances.
  In cross-checks
  we add more freedom to the $P$ and $D$ waves by replacing the $K^*(1680)$ and $K^*_2(1430)$,
  respectively, Breit-Wigner shapes with the binned amplitudes,
  as illustrated in Figs.~\ref{fig:Pwave} and \ref{fig:Dwave}. 
  We did not find improvement in the fit quality or any
  significant deviation of the binned wave from the analytic function of
  the isobar model.
  The systematic uncertainties described above are larger than variations  
  caused by the possible model dependence of the $P$ and $D$ waves.

  In model~I2 and QMIPWA we find that the $I=2$ $\pi^+\pi^+$ 
  $S$ wave amplitude is not consistent with a constant term as seen in Fig.~\ref{fig:pipiSwave}. 
  Its behavior is well modeled by the analytic function
  [Eq.~(\ref{eqn:I2pipiSwave})]; the binned wave also describes the behavior consistently.

\subsection{Comparison of fit models}
\label{sec:comparison_of_fit_models}

Note that in model~C, model~I2, and QMIPWA 
(Tables~\ref{tab:model_parameters}-\ref{tab:Isobar_I2_vs_QMIPWA_I2})
the magnitudes and phases for the resonance contributions
$K^*(892)$ (by construction), $K_0^*(1430)$, $K_2^*(1430)$, and
$K^*(1680)$ are consistent.  On the other hand, the fit fractions for the
resonances differ significantly among the models. The differences in fit
fractions arise from our models of the $K\pi$ $S$ wave 
and the $I=2$ $\pi^+\pi^+$ $S$ wave when it is included.
Indeed, in model~C the $K\pi$ $S$ wave amplitude is represented by the $NR$
term and a broad Breit-Wigner resonance for the $\kappa$. They have small magnitudes,
and they interfere constructively.  
This leads to small fit fractions
and the sum of all fit fractions is less than 100\%.
In model~I2 the $K\pi$ $S$ wave amplitude is represented by the $NR$ term
and the complex pole for the $\kappa$. 
The $I=2$ $\pi^+\pi^+$ $S$ wave is also included in this model.
These amplitudes are large, and they interfere destructively. This leads to
large fit fractions, and the sum
of all fit fractions is greater than 100\%.
Both models describe the total $K\pi$ $S$ wave 
with complex functions that qualitatively
show very similar behavior. 
However, the fit fractions,
especially for the components of the $K\pi$ $S$ wave,
strongly depend on the assumed composition of this amplitude.

In contrast the QMIPWA represents the $K\pi$ $S$ wave amplitude
as a single function.
We make no assumptions about the composition of the $K\pi$ $S$ wave in this
case. It also includes the I=2 $\pi^+\pi^+$ $S$ wave,
which plays a key role in improving the fit quality.
We feel this approach is the most reliable presentation of
results for the fit fractions for resonance and $S$ wave contributions
to the $D^+ \to K^-\pi^+\pi^+$ decay.

\section{Summary}
\label{sec:summary}

We describe a partial wave analysis of the $D^+ \to K^- \pi^+ \pi^+$ 
events on the Dalitz plot.
We use the CLEO-c data set of 572~pb$^{-1}$ of $e^+e^-$ collisions
accumulated at the $\psi(3770)$, which corresponds to 
a sample of $1.6 \times 10^6$ $D^+D^-$ pairs produced in the process 
$e^+e^- \to \psi(3770) \to D^+D^-$.
We select 140793 candidate events for
the Dalitz plot with a small background of 1.1\%.
We compare our results with the best previous measurements from E791~\cite{E791-2002}
using the isobar model. 
Our results agree with the E791 measurement, 
as shown in Table~\ref{tab:modelC} for their best model.
The fit quality can be improved if we add the $I=2$ $\pi^+\pi^+$ $S$ wave
contribution, as presented in Table~\ref{tab:Isobar_I2_vs_QMIPWA_I2}.

We apply a model-independent approach, developed in Ref.~\cite{E791_Kpipi}, 
to measure the magnitude and phase of the $K\pi$ $S$ wave in the 
invariant mass range from threshold to the maximum value in this decay.
In contrast to E791, we have measured the $K\pi$ $S$ wave without
factorization of the form factor for scalar resonances. 
Our results on the $K\pi$ $S$ wave phase and
magnitude measurement are presented in Table~\ref{tab:results_Swave_with_I2}
and in Fig.~\ref{fig:Swave_QMIPWA_I2}. 
The accuracy of the $K\pi$ $S$ wave measurement is improved,
compared to the only previous measurement~\cite{E791_Kpipi}. 
We find that the total observed $S$ wave magnitude 
in the $D^+ \to K^- \pi^+ \pi^+$ decay is essentially constant
from $K\pi$ production threshold to 1.4~GeV/$c^2$.
The phase shows smooth variation from 
--80$^\circ$ to 40$^\circ$ in the same range. 
At higher invariant mass $m_{K\pi}>1.4$~GeV/$c^2$,
the $S$ wave behavior is dominated by the $K_0^*(1430)$ resonance.
We find that the $P$ wave contribution is dominated by
$K^*(892)$ and $K^*(1680)$ Breit-Wigner resonances, and 
the $D$ wave has only a contribution from $K_2^*(1430)$.
Using binning techniques, we find no significant deviation
of the $P$ and $D$ waves from the isobar model, as demonstrated in 
Figs.~\ref{fig:Pwave} and \ref{fig:Dwave}.

In the model-independent approach 
for the $I=2$ $\pi^+\pi^+$ $S$ wave we obtain the binned amplitude
parameters, listed in Table~\ref{tab:results_I2_pipi_Swave}
and shown in Fig.~\ref{fig:pipiSwave},
which are consistent with the analytic form of this wave.
We find that the $I=2$ $\pi^+\pi^+$ $S$ wave has a nonuniform 
variation in the amplitude across the $m(\pi^+\pi^+)$ kinematic range 
with a fit fraction of 10\%--15\%.
As expected, the measured amplitude behavior 
and the fit fraction of the $K\pi$ $S$ wave 
changes slightly with the addition of the $I=2$ $\pi^+\pi^+$ $S$ wave.
The addition of the $I=2$ $\pi^+\pi^+$ $S$ wave to either the isobar 
model or the model-independent partial wave approach is the key piece that
gives good agreement with the data in both cases.

\section*{Acknowledgments}

We gratefully acknowledge the effort of the CESR staff
in providing us with excellent luminosity and running conditions.
D.~C.-H. and A.~R. thank the A.P.~Sloan Foundation.
This work was supported by the National Science Foundation,
the U.S. Department of Energy,
the Natural Sciences and Engineering Research Council of Canada, and
the U.K. Science and Technology Facilities Council.

\appendix

\section{Appendix: Kinematic variables and angular distributions} 
\label{app:angular}

Following previous CLEO analyses (see, for example, Ref.~\cite{Tim})
we use the angular distributions
[Eqs.~(\ref{eqn:angular_scalar})--(\ref{eqn:angular_tensor})]
obtained from a covariant-tensor formalism.
The E791 form \cite{E791_Kpipi} was applied
to the orbital momentum partial wave decomposition using
the term $(-2 {\rm P}_a {\rm P}_c)^L {\mathcal P}_L(\cos\theta)$ in stead of 
$\Omega_L(s,t)$ in Eq.~(\ref{eqn:PWA_amplitude}).
Here ${\mathcal P}_L(\cos\theta)$ is a Legendre polynomial, 
${\rm P}_a$, ${\rm P}_c$, and $\theta$ are the momenta and the angle between particles $a$ and $c$ 
in the resonance $R$ rest frame, respectively.
In this section we show that both approaches are equivalent up to constant coefficients.
As a by-product we also obtain expressions for the decay momenta used in the Blatt-Weisskopf
form factors [Eqs.~(\ref{eqn:formf_scalar})--(\ref{eqn:Gaussian_form_factor})].

The kinematic variables in the decay under study, schematically shown in 
Fig.~\ref{fig:quasitwobody},
can be expressed in terms of invariant variables (masses and invariant masses)
in the decaying particle $d$ ($D^+$ meson) or resonance $R$ rest frames. 
The energy and momentum of particle $a$ in the resonance $R$ or $(ab)$ rest frame
can be obtained from the 4-momentum balance equation
$p_b = p_R - p_a$, where $p_R = (m_{ab},\vec{0})$ and $p_a=(E_a,{\mathbf P}_a)$.
Then, $p_b^2 = m_b^2 =  m_{ab}^2 + m_a^2 - 2m_{ab}E_a$, giving the energy
\begin{equation}
\label{eqn:Ea}
        E_a = \frac{m_{ab}^2 + m_a^2 - m_b^2}{2m_{ab}},
\end{equation}
and relevant momentum squared
\begin{equation}
\label{eqn:Pa2}
   {\rm P}_a^2=E_a^2-m_a^2
   =\frac{1}{4}\bigg(m^2_{ab} - 2 m^2_a - 2 m^2_b  + \frac{(m^2_a - m^2_b)^2}{m^2_{ab}} \bigg).
\end{equation}
The energy and momentum of the particle $c$ in the resonance $R$ rest frame
can be obtained from the particle $d$ invariant mass squared, $m_d^2 = (p_c + p_R)^2$.
In the resonance rest frame 
$p_c=(E_c,{\mathbf P}_c)$ and $p_R = (m_{ab},\vec{0})$, and therefore
$m_d^2 = m_c^2 + m_{ab}^2 + 2 m_{ab} E_c$, from which we find its energy 
\begin{equation}
\label{eqn:Ec}
        E_c = \frac{m_d^2 - m_{ab}^2 - m_c^2}{2m_{ab}},
\end{equation}
and the associated momentum squared
\begin{equation}
\label{eqn:Pc2}
  {\rm P}_c^2=E_c^2-m_c^2
  =\frac{1}{4}\bigg(m^2_{ab} - 2 m^2_d - 2 m^2_c  + \frac{(m^2_d - m^2_c)^2}{m^2_{ab}} \bigg).
\end{equation}

The energy and momentum of the same particle $c$ in the $D$ meson rest frame,
denoted here by the asterisk, have different expressions:
\begin{equation}
\label{eqn:Ec_and_Pc2_in_D_com}
        E_c^* = \frac{m_d^2 - m_{ab}^2 + m_c^2}{2m_d}, ~~~~
        {\rm P}_c^{*2}=E_c^{*2}-m_c^2.
\end{equation}

The angular distributions in the E791 \cite{E791_Kpipi} analysis are defined by the Legendre polynomials ${\cal P}_L(\cos\theta)$, 
\begin{equation}
   \label{eqn:Legendre_polynomials}
    {\mathcal P}_0(x) = 1, ~~
    {\mathcal P}_1(x) = x, ~~
    {\mathcal P}_2(x) = \frac{3x^2  - 1}{2}, ~~
    {\mathcal P}_3(x) = \frac{5x^3  - 3x}{2}, 
     \ldots ,
\end{equation}
where $\theta$ is
the angle between particles $a$ and $c$
in the resonance $R$ rest frame. 
This $\cos\theta$
can be expressed through the known energies and momenta of particles $a$ and $c$
and their measured invariant mass squared
$m_{ac}^2 = (p_a + p_c)^2 = m_a^2 + m_c^2 + 2 E_a E_c - 2 {\rm P}_a {\rm P}_c \cos\theta$, so 
\begin{equation}
\label{eqn:cos_theta_1}
        \cos\theta = \frac{m_a^2 + m_c^2 + 2 E_a E_c - m_{ac}^2}{2 {\rm P}_a {\rm P}_c}.
\end{equation}

Substituting $E_a$ and $E_c$ from Eqs.~(\ref{eqn:Ea}) and (\ref{eqn:Ec})
in the numerator of Eq.~(\ref{eqn:cos_theta_1}) we get
\begin{equation}
\label{eqn:cos_theta_2}
        \cos\theta = \frac{1}{4 {\rm P}_a {\rm P}_c}
                     \bigg( m^2_{bc} - m^2_{ac} + \frac{(m^2_d - m^2_c)(m^2_a - m^2_b)}{m^2_{ab}} \bigg).
\end{equation}

Note that these angular distributions are equivalent 
to formulas used in 
Eqs.~(\ref{eqn:angular_scalar})--(\ref{eqn:angular_tensor})
up to constant factors. Indeed, comparing the expressions for
$(-2 {\rm P}_a {\rm P}_c)^L  {\cal P}_L(\cos\theta)$ 
from Ref.~\cite{E791_Kpipi} with $\Omega_{L}$ we get

\begin{eqnarray}
   \label{eqn:Angular_CLEO}
     (-2 {\rm P}_a {\rm P}_c)^0 {\mathcal P}_0(\cos\theta) = & 1  
                                                   & = ~~ \Omega_{L=0}, \\
     (-2 {\rm P}_a {\rm P}_c)^1 {\mathcal P}_1(\cos\theta) = & -2 {\rm P}_a {\rm P}_c \cos\theta 
                                                   & = \frac{1}{2}   \Omega_{L=1}, \\
     (-2 {\rm P}_a {\rm P}_c)^2 {\mathcal P}_2(\cos\theta) = & 4 ({\rm P}_a {\rm P}_c)^2   \frac{3\cos^2\theta-1}{2}
                                                   & = \frac{3}{8}   \Omega_{L=2}.
\end{eqnarray}


\begin{thebibliography}{99}
\bibitem{PDG-2006}  W.-M.~Yao {\it et al.},  Journal of Physics G {\bf 33}, 1 (2006).  
\bibitem{MARK-III-1987} J.~Adler  {\it et al.} (MARK III Collaboration), Phys. Lett. B {\bf 196}, 107 (1987).
\bibitem{NA14-1991} M.P.~Alvarez  {\it et al.} (NA14/2 Collaboration), Z. Phys. C {\bf 50}, 11 (1991).
\bibitem{E691-1993} J.C.~Anjos    {\it et al.} (E691 Collaboration), Phys. Rev.  D {\bf 48},  56 (1993).
\bibitem{E687-1994} P.L.~Frabetti {\it et al.} (E687 Collaboration), Phys. Lett. B {\bf 331}, 217 (1994).
\bibitem{E791-2002}  E.M.~Aitala  {\it et al.} (E791 Collaboration), Phys. Rev. Lett. {\bf 89}, 121801 (2002).
\bibitem{E791_Kpipi} E.M.~Aitala  {\it et al.} (E791 Collaboration), Phys. Rev. D {\bf 73}, 032004 (2006).
\bibitem{Dalitz} R.H.~Dalitz, Philos. Mag. {\bf 44}, 1068 (1953).
\bibitem{Bugg-2005} D.~Bugg, hep-ex/0510014.
\bibitem{Bugg-2006} D.~Bugg, Phys. Lett. B {\bf 632}, 471 (2006).
\bibitem{Oller-2005} J.A.~Oller, Phys. Rev. D {\bf 71}, 054030 (2005).
\bibitem{Hoodland} W.~Hoogland {\it et al.}, Nucl. Phys. B {\bf 126}, 109 (1977);
                   N.B.~Durusoy {\it et al.}, Phys. Lett. {\bf 45B}, 517 (1973);
                   B.S.~Zou {\it et al.}, published in Hadron 2003 Proceedings,
                                               AIP Conf. Proc. 717:347 (2004) [hep-ph/0405118].
\bibitem{CLEO-c} G.~Viehhauser, Nucl. Instrum. Methods A {\bf 462}, 146 (2001);
        D.~Peterson {\it et al.}, Nucl. Instrum. Methods Phys. Res., Sect. A {\bf 478}, 142 (2002);
        Y.~Kubota {\it et al.}, Nucl. Instrum. Methods Phys. Res., Sect. A {\bf 320}, 66 (1992);
        R.A.~Briere {\it et al.} (CESR-c and CLEO-c Taskforces, CLEO-c Collaboration),
        Cornell University, LEPP Report No. CLNS 01/1742 (2001) (unpublished).
\bibitem{HadronicBF} S.~Dobbs {\it et al.} (CLEO Collaboration), Phys. Rev. D {\bf 76}, 112001 (2007).
\bibitem{ARGUS_fun} H.~Albrecht {\it et al.}, Phys. Lett. B {\bf 229}, 304 (1989).
\bibitem{D3pi}   G.~Bonvicini {\it et al.} (CLEO Collaboration), Phys. Rev. D {\bf 76}, 012001 (2007).
\bibitem{EVTGEN} D.J.~Lange, Nucl. Instrum. Methods Phys. Res., Sect. A {\bf 462}, 152 (2001).
\bibitem{Tim}    S.~Kopp {\it et al.} (CLEO Collaboration), Phys. Rev. D {\bf 63}, 092001 (2001).
\bibitem{Blatt-Weisskopf} J.M.~Blatt and V.F.~Weisskopf, {\it Theoretical Nuclear Physics}, Wiley, New York, 1951, p. 361.
\bibitem{Tornquist} N.A.~Tornqvist, Z. Phys. C {\bf 68}, 647 (1995).
\bibitem{Achasov_PRD67_2003} N.N.~Achasov and G.N.~Shestakov, Phys. Rev. D {\bf 67}, 114018 (2003).
\bibitem{PDG-2000}  D.E.~Groom {\it et al.}, The European Physical Journal C {\bf 15}, 1 (2000).
\bibitem{FOCUS-2007} J.M.~Link {\it et al.} (FOCUS Collaboration), 
                                            Phys. Lett. B {\bf 653}, 1 (2007). 
\bibitem{FOCUS-Kst-width} J.M.~Link {\it et al.} (FOCUS Collaboration), 
                                            Phys. Lett. B {\bf 621}, 72 (2005).
\bibitem{Belle-Kst-width} D.~Epifanov {\it et al.} (Belle Collaboration), 
                                            Phys. Lett. B {\bf 654}, 65 (2007).

\end{thebibliography}
\end{document}